\documentstyle[aas2pp4,psfig]{article} 
\def\etal{{\it et al. }}
\def\Msun{M$_{\odot}$}
\def\rcolhead#1{\multicolumn{1}{r}{#1}}
\newcommand{\BC}{Bruzual and Charlot}

\newcommand{\degree}{$^{\rm o}$}
\newcommand{\speed}[1]{$#1 \rm \, km \, s^{-1}$}
\newcommand{\speedpm}[2]{#1$\pm$#2$\rm\,km \, s^{-1}$}
\newcommand{\speedas}[3]{#1$^{+#2}_{-#3} \rm\,km \, s^{-1}$}
\newcommand{\hubunits}[1]{#1 $\rm \! km \, s^{-1} \, Mpc^{-1}$}
\newcommand{\Ho}{H$_0$}
\newcommand{\abmagn}[2]{$\rm M_{#1}\!\!=\!\!#2$}
\newcommand{\magn}[2]{$\rm #1\!=\!#2$}
\newcommand{\dismod}[2]{$\rm (m\!-\!M)_{#1}\!=\!#2$}
\newcommand{\absorb}[1]{A$\rm _#1$}
\newcommand{\absorbval}[2]{$\rm A_#1 \!=\! #2 $}
\newcommand{\about}[1]{$\sim \,$#1}
\newcommand{\Sn}{$S_N$}

\newcommand{\valpm}[2]{#1$\pm$#2}

\newcommand{\color}[2]{($\rm #1\!-\!#2$)}
\newcommand{\excessval}[3]{$\rm E_{(#1\!-\!#2)}\!=\!#3$}
\newcommand{\app}{$\!{\rm \AA/pixel}$}
\newcommand{\Halpha}{H$\alpha$}
\newcommand{\Hbeta}{H$\beta$}
\newcommand{\Hgamma}{H$\gamma$}
\newcommand{\Hdelta}{H$\delta$}
\newcommand{\Mdot}[1]{$\rm #1 \, M_{\odot} \, yr^{-1}$}

\newcommand{\Mdotas}[3]{$\rm #1^{+#2}_{-#3} \, M_{\odot} \, yr^{-1}$}\newcommand{\agerangeM}[2]{$\rm #1\!-\!#2 \, Myr$}
\newcommand{\agerangeG}[2]{$\rm #1\!-\!#2 \, Gyr$}
\newcommand{\Mrange}[3]{$\rm #1\!-\!#2 \, M_{\odot} \,\,$}
\newcommand{\seqorder}[3]{#1$<$#2$<$#3}
\newcommand{\gc}{globular cluster}
\newcommand{\gcs}{globular clusters}
\newcommand{\pgcc}{proto--globular cluster candidate}
\newcommand{\pgccs}{proto--globular cluster candidates}
\newcommand{\GCS}{globular cluster system}

\begin{document}
\input{psfig.sty}
\title{Keck Spectroscopy of Candidate Proto--globular Clusters in NGC 1275$^1$}

\author{Jean P. Brodie}
\affil{Lick Observatory, University of California, Santa Cruz, CA 95064}
\affil{Electronic mail: brodie@ucolick.org}

\author{Linda L. Schroder}
\affil{Lick Observatory, University of California, Santa Cruz, CA 95064}
\affil{Electronic mail: linda@ucolick.org}

\author{John P. Huchra}
\affil{Harvard-Smithsonian Center for Astrophysics, 60 Garden St, Cambridge MA 01238}
\affil{Electronic mail: huchra@cfa.harvard.edu}

\author{Andrew C. Phillips}
\affil{Lick Observatory, University of California, Santa Cruz, CA 95064}
\affil{Electronic mail: phillips@ucolick.org}

\author{Markus Kissler-Patig}
\affil{Lick Observatory, University of California, Santa Cruz, CA 95064}
\affil{Electronic mail: mkissler@ucolick.org}

\author{Duncan A. Forbes}
\affil{School of Physics and Astronomy, University of Birmingham, Birmingham B15 2TT, England}
\affil{Electronic mail: forbes@star.sr.bham.ac.uk}

\altaffiltext{1}{Based on observations obtained at the W. M. Keck Observatory,
which is operated jointly by the California Institute of Technology and
the University of California.}

\begin{abstract}

Keck spectroscopy of 5 \pgccs~in NGC 1275 has
been combined with HST WFPC2 photometry to explore the nature and origin of
these objects and discriminate between merger and cooling flow scenarios for
\gc~formation.

The objects we have studied are not HII regions, but rather star clusters,
yet their integrated spectral properties do not resemble young or
intermediate age Magellanic Cloud clusters or Milky Way open clusters.  The
clusters' Balmer absorption appears to be too strong to be consistent with
any of the standard \BC~evolutionary models at any metallicity. If the
\BC~models are adopted, an IMF which is skewed to high masses provides a
better fit to the \pgccs~ data.  A truncated IMF with a mass range of
\Mrange{2}{3}~reproduces the observed Balmer equivalent widths and colors
at \about{450} Myr.

Formation in a {\it continuous} cooling flow appears to be ruled out since
the age of the clusters is much larger than the cooling time, the spatial
scale of the clusters is much smaller than the cooling flow radius, and the
deduced star formation rate in the cooling flow favors a steep rather than
a flat IMF.  A merger would have to produce clusters only in the central
few kpc, presumably from gas in the merging galaxies which was channeled
rapidly to the center.  Widespread shocks in merging galaxies cannot have
produced these clusters.

If these objects are confirmed to have a relatively flat, or truncated, IMF
it is unclear whether or not they will evolve into objects we would regard as
{\it bona fide} \gcs.

\end{abstract}


\section{Introduction}

The discovery by Holtzman \etal (1992) of \pgccs~in 
NGC 1275, the peculiar cD galaxy at the center of the Perseus cluster,
was regarded by many as an important piece of evidence in favor of the
merger model for elliptical galaxy formation (Schweizer 1987, Ashman \&
Zepf 1992).  The Ashman \& Zepf model implies that \gcs~are
formed with high efficiency during a major merger of two gas--rich spiral
galaxies.  It has also been suggested (e.g. Fabian \etal1984) that \gcs~can
form efficiently in cooling flows.  NGC 1275 has one
of the largest known cooling flows (Fabian \etal1981, Allen \& Fabian 1997)
and it has a Seyfert--type nucleus with strong \Halpha~emission, as well as a
large molecular and neutral hydrogen content. A number of features including
its irregular shape may be indicative of a merger
history (e.g. Zepf \etal1995). 

Minkowski (1957) discovered the existence of two velocity systems in the
galaxy spectrum.  The low--velocity (LV) system, comprising filaments of
ionized gas whose radial velocity is similar to that of the main body of
NGC 1275 (i.e.~\speedpm{5264}{11}, Strauss \etal1992), is generally thought
to be a product of the cooling flow.  Lazareff \etal(1989) detected
significant amounts (\about{6 x 10$^9$} \Msun) of CO associated with the
\Halpha~filaments, perhaps revealing the source of raw materials (molecular
gas) from which stars/clusters might form. The association of the molecular
gas with the ionized filaments was inferred from the CO and
\Halpha~spectral line parameters (central velocity, line width and
integrated intensity) which displayed similar changes between the two
locations corresponding to the radio beam positions of the CO study.

The CO gas was found to have such a low velocity dispersion that it should
collapse into stars before it virializes in the gravitational potential of
the galaxy.  According to Lazareff \etal(1989), this implies cooling times of $<$ 3 x 10$^7$ yr which, with 6 x
10$^9$ \Msun~of material, leads to a star formation rate of $>$ \Mdot{200}.
Allen \& Fabian (1997) derive an even larger flow rate (\Mdotas{456}{16}{9})
from X--ray emission measurements.
Far UV colors (Smith \etal1992) indicate that there are essentially no high
mass stars
($>$ 5 \Msun) in NGC 1275 so a continuous cooling flow would have to produce
stars with a steep IMF (skewed to low masses) or else star formation must
have stopped at least 80 Myr ago.  The 2--dimensional spectroscopy of
Ferruit \& P\'{e}contal (1994) indicates that the \pgccs~are associated
with dense ionized gas regions. For example, the morphology of their [NII]
isophote contour map corresponds quite closely to the distribution of the
clusters. On the basis of their kinematics, these dense gas regions are
thought to be associated with LV infalling filaments. There appears, then,
to be evidence that the clusters, the LV ionized gas filaments and the
CO gas are all associated with each other.

There is also a system of giant HII clouds with a velocity of
\about{\speed{3000}}
relative to the main body of NGC 1275, referred to in the literature
as the high--velocity (HV) system.  The physical relationship of the HV
system to the LV system is a matter of some uncertainty.  Multi--wavelength
observational evidence (summarized by Kaisler \etal1996) seems to indicate
that this HV system is an infalling galaxy, still in front of NGC 1275.
However, N{\o}rgaard-Nielsen \etal (1993) argue that the major regions of
extinction, which are probably associated with the HV system, are situated
well within the main body of the galaxy.  They conclude that the HV
system is approximately halfway through the main body of the galaxy
and/or the LV system.  In any
case, unless the \pgccs~are very young ($<$40 Myr), it is unlikely that the HV
system is associated with their formation.  Even if the
HV system has proceeded halfway through the LV system it will have had only
\about{10$^6$} years to interact, given the high relative velocity of the two
systems (Richer \etal1993).
 
It is well known (e.g.~Harris, 1991) that the specific frequency of
\gcs~(number of clusters per unit galaxy light, \Sn) is larger in giant
ellipticals than in spiral galaxies and it is roughly three times larger
for cD galaxies residing in rich clusters of galaxies than for ``normal''
giant ellipticals.  A number of suggestions have been made to explain these
high specific frequencies [see Forbes, Brodie \& Grillmair (1997) for a
summary] and many of these can be tested by reference to NGC 1275. In
particular, if \gc~formation in cooling flows (Fabian \etal1984) and/or
mergers (Ashman \& Zepf, 1992) are important factors in determining \Sn,
the evidence should be present in NGC 1275.

West \etal(1995) were the first to suggest that the number of ``excess''
\gcs~above a ``normal'' \Sn~correlates with the galaxy cluster's X--ray
temperature (T$_X$ $\propto$ M$_{cluster}$). They contend that if a galaxy
is at, or close to, the center of a rich cluster of galaxies, a high
\Sn~globular cluster system should be found.  Both a central location and
the cluster richness are necessary conditions for a high value of \Sn.  The
most recent work by Blakeslee (1996), using a homogeneous data base
comprising 5 cD and 14 gE galaxies, indicates that \Sn~is indeed strongly
correlated with cluster density as measured by velocity dispersion, the
temperature and luminosity of the X--ray emitting gas and, to a lesser
extent, the local density of bright galaxies.  Blakeslee's view is that
bright central galaxies with high \Sn~values, which are usually but not
always cDs, are not anomalous in the sense that they have too many \gcs~for
their luminosity.  Rather, these galaxies are underluminous for their
central position, and the number of \gcs~accurately reflects the local
galaxy cluster mass density. It now appears that cluster giant elliptical
and cD galaxies cover a continuum of \Sn~values and these values depend on
properties of the local environment.  The fact that there is never more
than one massive, high \Sn~galaxy per cluster is consistent with the West
\etal and Blakeslee results.

NGC 1275 is located at the center of the X--ray emission in Perseus (Allen
\etal1992).  The central X--ray temperature and the velocity dispersion of
the cluster are 7.5 keV (Arnaud \etal1994) and \speedas{1277}{95}{78}
(Zabludoff \etal 1990) respectively.  Applying these values to the
relations illustrated in Blakeslee (1996, Figs. 5.32 and 5.27) leads to
predicted \Sn~values for NGC 1275 of approximately 9 and 11 respectively,
albeit with large uncertainties.
  
Overall, then, a high specific frequency is to be expected for the old
\GCS~of NGC 1275.  The ``old'' \gc~population has been studied by Kaisler
\etal(1996) who found a \Sn~value of \valpm{4.4}{1.2}, much lower than
expected for a central cD galaxy in a rich cluster
(\about{\seqorder{11}{$S_N$}{15}}) and typical of values found for
``normal'' giant ellipticals. However, the value they derive is heavily
dependent on the values of apparent magnitude and distance they adopt for
this galaxy.  Kaisler \etal adopted \magn{V}{11.59} (from RC2) and deduce a
distance modulus \dismod{V}{34.9} assuming \Ho=\hubunits{75} and
\absorbval{V}{0.6}. They then derived \abmagn{V}{-23.3} and corrected this
upward by 15\% to \abmagn{V}{-23.14} to take account of blue light from
starburst activity. We have reworked the numbers adopting \magn{V}{11.88}
(from RC3) and using a slightly different velocity, {\it v} = \speed{5264}
instead of {\it v} = \speed{5200}.  We calculate \dismod{V}{34.83} with the
same assumptions as Kaisler \etal, i.e.~\Ho~=75 and \absorbval{V}{0.6}.  We
derive \abmagn{V}{-22.87}, or \abmagn{V}{-22.72} with the same 15\%
correction (arguably an underestimate but we adopt it for consistency with
the Kaisler \etal calculations).  Using the total number of \gcs~inferred
from Kaisler \etal(i.e.~\about{7900}), this absolute V magnitude results in
a specific frequency of 6.5 rather than the 4.4 quoted by Kaisler \etal
Furthermore, the total number of \gcs~is quite poorly constrained by the
Kaisler \etal data, which samples only a small fraction ($<$4\%) of the
globular cluster luminosity function. They note that the last data point at
the faint end of their distribution (still much brighter than the peak of
the luminosity function) is surprisingly low. The total number of
\gcs~could easily be in error by 30\% or more, given the large
extrapolation needed to make this estimate.  A 30\% increase would easily
bring the specific frequency in line with the expected value of
\about{10}. In fact, Carlson \etal (1998) report S$_N \sim 10$ for the old
\gcs~in NGC 1275. Hence it may not be necessary to invoke the formation of
large numbers of ~\gcs~in the more recent past to produce the expected high
specific frequency.

The initial discovery of \pgccs~by
Holtzman \etal(1992) from HST WFPC1 data was later confirmed
in the ground--based imaging study of Richer \etal(1993). Holtzman
\etal claimed that the photometric properties of the young clusters
are consistent with their evolution to objects resembling Milky
Way \gcs. From the photometric data, Holtzman \etal favored
their origin in a merger event, whereas Richer \etal suggested formation
in a cooling flow.
In principle, the age spread of the clusters can discriminate
between these different formation scenarios.
Clusters formed in a recent merger should
all be young, with ages no greater than perhaps a few hundred million years.
Repeated mergers will,
of course, confuse this simple picture.
Continuous formation in a cooling flow would lead to a spread in ages
with some clusters possibly older than a billion years (Faber 1993). 

Although recent HST observations have revealed \pgccs~in
several other ``merger'' galaxies (e.g. Whitmore \etal1993,
Whitmore \& Schweizer 1995, Holtzman \etal1996), none have been found in a
cooling flow galaxy (Holtzman \etal1996).  
Attempts to deduce the age spread of NGC 1275 \pgccs~from the observed
color spread (Holtzman \etal1992, Richer \etal1993) have been undermined by
problems with the photometry (Richer \etal1993, Faber 1993).  The absence
of a ``fading vector'' (with fainter clusters being redder) in their
data was one
of the reasons (along with a narrow color spread, later disputed by Richer
\etal1993) that led Holtzman \etal to suggest a narrow age spread in the
\pgcc~system. However, the simulations of Faber
(1993) indicate that none of the probable formation scenarios
is likely to produce an observable
``fading vector''. Interestingly, though, Faber's simulations predict a
significant difference between the merger and the cooling flow pictures in
the number of faint clusters expected to be present at magnitudes just
below the WFPC1 detection limit. WFPC2 data will be relevant to
this issue (Holtzman et al.~1998).

The next step in understanding the NGC 1275 system is to verify that the
\pgccs~are indeed clusters rather than HII regions. Once their cluster
status is confirmed, determining their ages and radial velocities will help
constrain the formation models.  Metallicity is unlikely to be a useful
diagnostic since, for very young objects (younger
than a few Gyr), little can be learned about metallicity from either
optical colors or moderate resolution spectra.

Velocity information can also provide clues about the origins of these
clusters. It may reveal bulk flows, should indicate directzly whether
these objects are associated with the high or low velocity gaseous systems
and whether the clusters are moving with the main body of the galaxy or
with the local flow of the surrounding gas.

Here we present the results of spectroscopy with the Keck I telescope of 8
of the \pgccs. In section 2 we discuss the
observations and data reduction, including a detailed discussion of the
critically important background--subtraction techniques. In section 3 we
discuss new photometric data from HST and compare them with previous estimates
from HST and ground--based telescopes. In section 4 we measure equivalent
widths in the cluster spectra and compare the clusters with spectroscopic
standards. In section 5 we present the radial velocity information. In
section 6 we estimate ages and set constraints on the IMF using
stellar evolutionary models and by comparison with spectra of
standard stars and star clusters.  Section 7 is a
discussion of the implications for \gc~formation models.
We end with a summary and conclusions.

\section{Observations and Data Reduction}

We observed 8 \pgccs~in NGC 1275 in 1994 Nov/Dec,
using the Low Resolution Imaging Spectrograph (LRIS, Oke \etal1995) on the
Keck I 10--m telescope.  The 600 l/mm grating, blazed at 7500 \AA, and a
slit width of 1\arcsec~provided a dispersion of 1.25~\app~and a
spectral resolution of 5--6 \AA.  Spectral coverage was 3900--6500~\AA.
Observational details are given in Table 1.

The galaxy spectrum of NGC 1275 consists of several components.  The LV
system, comprising the galaxy starlight and filamentary ionized gas,
contains strong Balmer emission with
broad [OIII] emission lines at a velocity close to the systemic velocity
of the galaxy proper (\speed{5264}). These emission lines become stronger
toward the nucleus while widespread patchy dust appears to be
significantly diminished very close to the nuclear region (See N{\o}rgaard-Nielson
\etal 1993 and Fig.~\ref{fig1}/Plate ??).  Superimposed on this LV material is the HV 
system with strong Balmer emission lines that are redshifted to a
velocity of \about{\speed{8000}}.  The emission lines for both systems vary
significantly on spatial scales that are comparable to the sizes of the
clusters themselves. This posed a serious challenge in determining an
appropriate background subtraction procedure for extracting the cluster
spectra.  Additional challenges in reducing the data arose from
instrumental limitations (discussed below) and the proximity of the cluster
candidates to the bright galaxy nucleus.  All the \pgccs~we
observed are within 5.3\arcsec~of the
galaxy center, corresponding to a distance of 1.9 $\rm h^{-1}\,kpc$ at the
distance of NGC 1275.  An individual cluster flux was $20\%$ of
the background flux for the brightest objects we observed and $4\%$ for the
faintest.

Because the grating blaze angle was quite red and detector was relatively
insensitive to blue light, the
system throughput at the blue end of the spectral range was quite poor.
In addition, the spectral flat fields, taken with internal
quartz--halogen lamps, contained spatially resolved artifacts due to
illumination and reflection off the wire--bond region of the CCD.  These
artifacts did not appear in the data. Their presence in the spectral flats,
combined with the low S/N in the blue, prevented flat--fielding the data on
the blue end of the spectrum.  Flat--fielding for the red end of the
spectrum was done in the usual way.  The 2--dimensional images were
rectified in both the x and y directions using the IRAF tasks IDENT,
REIDENT, FITCOORDS on spectra of comparison lamps and then the derived
transformation was applied to the science images using the IRAF task TRANSFORM.
The night sky background was subtracted using the IRAF task BACKGROUND.
What remained were 2--dimensional images containing the spectra of the LV
system, the HV system and the clusters.

The extraction procedure consisted of subtracting a representative galaxy
spectrum from each of the 2--dimensional images, scaled to match
the continuum level at the location of each cluster. 
Because of the multiple components and their variability on small spatial
scales, the underlying galaxy light profile was not
easily modeled from the 2--dimensional spectra alone.  First attempts to
determine the precise continuum level using only the LRIS spectral
images, containing the combined galaxy, gas and cluster spectra, proved
unsuccessful.

Ultimately, HST WFPC2 images of the region were used to model the galaxy
profile at the position angle of each observation.  The WFPC2 images
were rotated to match the orientation of each LRIS spectral
image and then convolved with a Gaussian function to mimic
the seeing of the Keck observations.  The WFPC2 images were
then rebinned to match the plate--scale of the Keck
spectral images.  Finally, columns across the ``slit'' on the WFPC2
image were inspected to determine which combination of columns best matched the combined
galaxy/gas/cluster light profile from the 2--dimensional Keck spectra.
After finding the best match, the entire procedure was exactly repeated on
the same WFPC2 image minus the clusters.  The
extracted light profiles from this second pass were used to determine the
galaxy continuum at the location of the cluster.  At most position angles
the galaxy light profile revealed itself to be more complex than would have
been possible to model using the 2--dimensional Keck spectra alone.  Not
surprisingly, this technique worked better for the higher S/N observations
than for the lower S/N ones.  For one position angle (158.1$^{\rm o}$) we were
not able to derive a satisfactory model of the galaxy light profile,
so no spectra were extracted from that set of observations.  Fortunately, the
clusters observed at that position angle, H5 and H6 (identification from
Holtzman \etal1992), were observed
at least at one other position angle.  Once the underlying continuum level was
determined, the major uncertainty (as far as spectral features are concerned)
was in selecting an accurate ``representative'' galaxy spectrum from
along the slit to subtract from the combined object+galaxy spectrum.  Once
the galaxy spectrum is removed a profile along the slit should
ideally show only the light profiles of the clusters.  In practice,
there generally remained some residual galaxy light, attributable
to an imperfect match between the representative galaxy
spectrum and the true galaxy spectrum at the location of the cluster.
Every effort was made to minimize this residual galaxy light, particularly
in the region of the spectrum where the Balmer lines are located.

With the general galaxy spectrum removed, the task of eliminating
the emission lines remained.  It was necessary to perform
these two operations (galaxy subtraction and emission line subtraction)
separately because spatial variations in the strengths and ratios of the
emission lines (for both the high and low velocity systems)
do not correlate with spatial variations in the underlying
galaxy spectrum/light profile.  The technique for removal of the emission
lines was simply to extract a sample of the emission lines from
either side of the cluster along the slit and find the weighted
average of the two samples that minimized the emission lines in the
cluster spectra themselves.  
Having extracted the cluster spectra, flux calibration proceeded in the
usual way. Corrections for atmospheric extinction and 
Galactic reddening (see Section 3) were made using the
IRAF/NOAO routines STANDARD, SENSFUNC and CALIBRATE.

Of the eight objects observed, two (H12 and H17) were
not detected above the noise and one (H3) was not
sufficiently resolved from its nearest, and much brighter, neighbor (H1)
for it to be extracted successfully.  We will not consider these objects
further.  The spectra of the remaining 5 objects are shown in
Fig.~\ref{fig2}.  Note that we have obtained a combined spectrum of H4 and
H9 which are resolvable as two separate objects in the HST images but are
not separable in the ground--based spectra.  As explained above, the
background contains several components each of which is spatially variable
and the scales and amplitudes of the variations are different.  Because the
background subtraction process could not perfectly remove each component
simultaneously, the continuum shapes of these spectra are not reliable and
emission lines may have been under--subtracted (note that the
background--subtraction process will not, in general, result in
over--subtraction).  Residual emission lines are particularly apparent in
the lower signal--to--noise spectra. It is impossible to be absolutely
certain that these emission lines do not belong to the objects themselves
but we believe they are artifacts, since the velocities of the
emission--line residuals are different from
the cluster velocities by several hundred km/s.
Moreover, these LV emission lines can be
removed by adjusting the background mix (at the price of introducing other
unacceptable features such as a negative continuum and/or absorption
features from the now oversubtracted HV system), indicating
that they belong to the background emission. 

The clusters' spectra do not resemble spectra of HII regions. All but the
lowest signal--to--noise spectrum, that of H2, contain absorption features
and a significant continuum component.  The emission lines are
demonstrably residuals from the background subtraction process.
In the case of H2 where the continuum component is very weak,
the presence of [NI] $\lambda 5199$ effectively rules out an HII region.
[NI] $\lambda 5199$ is distinctly present in the spectrum
of the nuclear region of the galaxy and H2 has the smallest radial
distance from the nucleus of our entire sample, suggesting that the [NI]
$\lambda 5199$ line in the spectrum of H2 is a residual from
the imperfect background subtraction.  We conclude that all 5
objects are {\it bona fide} clusters.

\section{Photometry}

There is some disagreement in the literature about the magnitudes and
colors of the \pgccs.
Richer \etal(1993) noted problems with the WFPC1 photometry of Holtzman
\etal(1992) and Faber (1993) indicated that the CFHT photometry of Richer
\etal(1993) must also be incorrect at some level.  The photometric studies
of N{\o}rgaard-Nielsen \etal(1993) 
and Ferruit \& P\'{e}contal (1994) do not agree particularly well
with either of the above nor with each other.
We elected to use
the WFPC2 images, taken from the HST archive, to derive our own B and R
magnitudes for the clusters. Using circular apertures and a
curve--of--growth type analysis we estimated total F450W and F702W
magnitudes. These were converted into standard Johnson B and R magnitudes
using the zeropoints and color transformations given in Holtzman
\etal(1995).  Our magnitudes and colors are presented in Table 2 along with
those of Holtzman \etal(1992), Richer \etal(1993), 
N{\o}rgaard-Nielsen \etal(1993) and Ferruit \& P\'{e}contal (1994).  
The Holtzman
\etal V magnitudes have been corrected for the acknowledged factor of 2
error, i.e.~they have been brightened by 0.75 mag (see Richer \etal1993).
Except for the H4+H9 clusters, 
the Ferruit \& P\'{e}contal R magnitudes are in excellent agreement
with our WFPC2 R values. 
All the entries in Table 2 have been corrected for 
foreground Galactic extinction corresponding to  \excessval{B}{V}{0.17} (Burstein \& Heiles 1984) and
A$_B$ = 0.71, A$_V$ = 0.54, A$_R$ =
0.40 and A$_I$ = 0.26.

\begin{figure}
\centerline{\psfig{figure=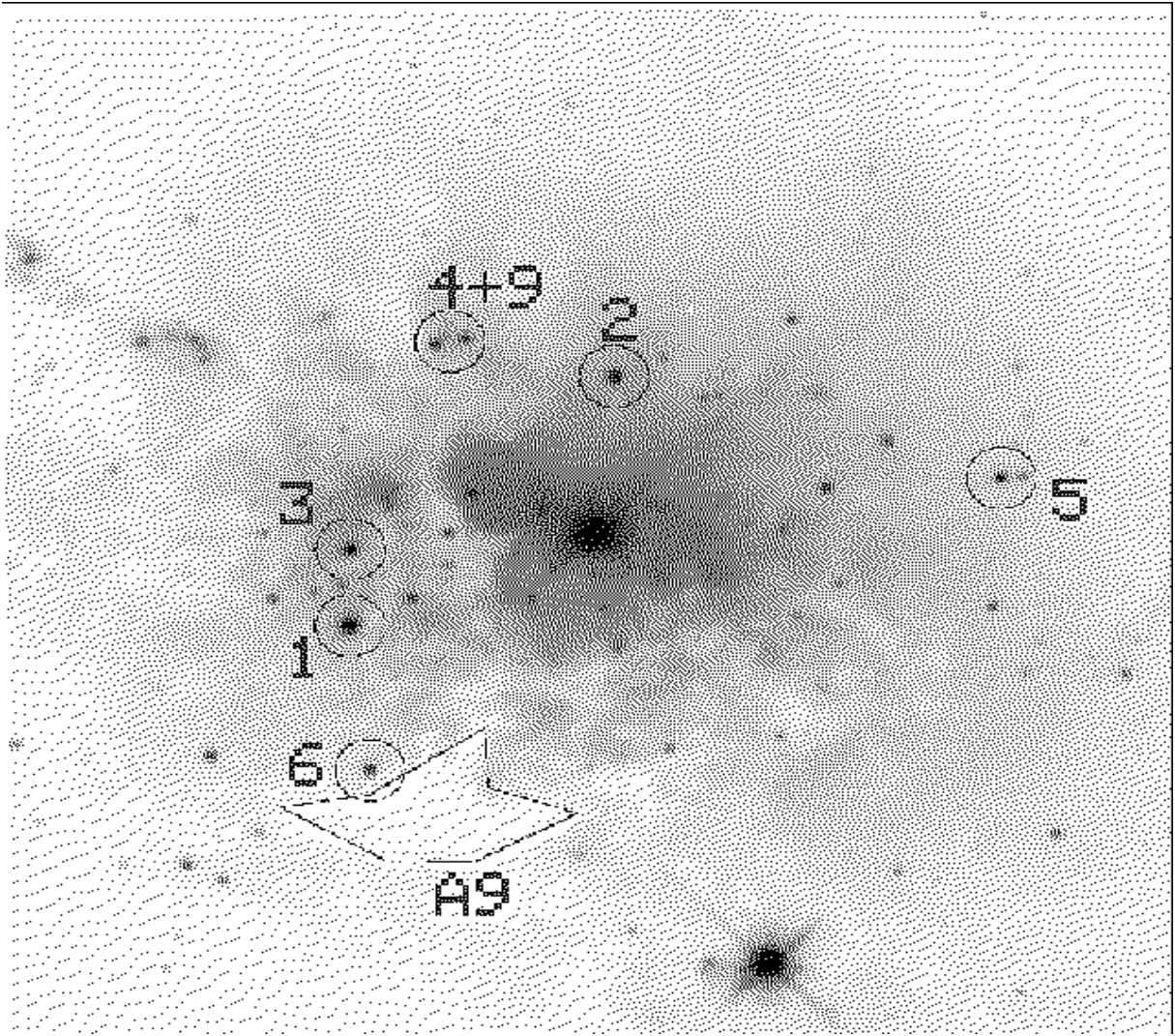,height=15cm,width=15cm,bbllx=22mm,bblly=38mm,bburx=199mm,bbury=238mm}}
\caption{\label{fig1} WFPC2 B image (F450W) of the inner 15\arcsec~of NGC 1275.
The 5 clusters for which we obtained spectra are circled.  H3, also
circled, was observed but its spectrum could not be
successfully extracted because of
its proximity to H1.  Also marked is the dust region A9 studied
by N{\o}rgaard-Nielsen \etal (1993).  The very bright object near
the bottom of the image is a Galactic F--star.  The image is oriented
such that North is 138\degree~counter--clockwise from the vertical.
East is 228\degree~clockwise from the vertical.}
\end{figure}

N{\o}rgaard--Nielson \etal(1993) modeled the dust extinction law and
estimated \absorb{V} and \absorb{I} for 9 regions (designated A1--A9)
in the galaxy lying between 4.7\arcsec~and 23\arcsec~from
the galaxy nucleus.  They found \seqorder{0.20}{$\rm A_V$}{0.81}
and \seqorder{0.19}{$\rm A_I$}{0.69}~for these regions.
Fig.~\ref{fig1}/Plate ?? is
a WFPC2 B image of the inner \about{15\arcsec} of NGC 1275.
The dust region 
A9, which appears on the WFPC2 PC image, is
marked on Fig.~\ref{fig1}/Plate ?? along with the 5 clusters whose
spectra are shown in Fig.~\ref{fig2}.  H3 is also marked but its
spectrum could not be extracted because of its proximity to
the much brighter H1.
Although none of the clusters appears by eye to be positioned directly in
a dust region as dense as A9
(\absorbval{V}{0.35}), variations in the underlying
background
brightness can easily disguise the contrast with dense dust clumps, 
particularly for regions closer to the galaxy center.
Interestingly, the three reddest clusters for which we
obtained spectra are clearly more affected by dust than the others.  H1 is
superposed on a dust filament of width comparable to the size of the cluster.
H6 and H4 are both located close to the apparent edges of dust regions.  Of
the bluer objects we observed, only H5 is in a region that is apparently
unaffected by patchy dust.

To quantify these impressions we created a B-R color image using the
HST data. For each cluster we measured the reddest color in the region
adjoining the cluster, and the bluest color in the galaxy at the same
radius as the cluster.  The difference between the reddest and bluest
color represents a rough value for the possible internal reddening for
each cluster, assuming that the bluest color represents the true
underlying color. These reddening estimates are included in Table 2.
The bluest regions of the galaxy may themselves be reddened, and
relatively uniformly distributed internal dust would not be detected
by this method. Therefore the clusters could still be bluer than the bluest
color limits we suggest.

The cluster apparent magnitude from the WFPC2 data can be used to derive a
mass estimate.  We used the population synthesis models of
Fritze-v.~Alvensleben \& Burkert (1995) to determine how much a cluster
will fade by the time it is $\sim$ 15 Gyr old. Assuming a present age of
$\sim$ 400 Myr (see section 6), a true distance modulus of 34.23 for NGC
1275, a V-band M/L ratio of 2 for the clusters (if the clusters do indeed
evolve into globular clusters they will have roughly this M/L ratio at
$\sim$ 15 Gyr) and Galactic extinction of \excessval{B}{V}{0.17}, we derive
a mass of \about{2 x 10$^7$} \Msun~for the brightest cluster, H1. This is
consistent with the value deduced by Holtzman \etal(1992).  In section 6 we
discuss the implications of color and reddening for age dating and
formation scenarios for the clusters.

We measured the sizes of all 7 {\it single} clusters on the WFPC2 images to
determine if the clusters are resolved.  At the distance of NGC 1275 a
typical Milky Way globular cluster (effective radius R$_{eff}$ $\sim$ 3 pc,
where R$_{eff}$ is the half--light radius,.
as defined as in Whitmore \etal
1993) would not be resolved, even with the \about{0.05\arcsec} resolution
of of the Planetary Camera (PC) on HST. At the distance of NGC 1275, one
pixel on the PC corresponds to $\sim$17 pc so the largest Milky Way
globular clusters R$_{eff}\sim$ 10 pc would be barely resolvable.  Using
the method developed and calibrated by Miller \etal (1997) in their study
of NGC 7252 young clusters, we have measured the FWHM and the magnitude
difference between 0.5 and 3.0 pixels radius apertures and we conclude that
none of these brightest NGC 1275 clusters are resolved.

\section{Spectral Types and Equivalent Widths}

The \pgccs~were compared with standard star spectra
covering a range of spectral types and luminosity classes. Ignoring the
overall continuum shape, which we have argued is unreliable because of the
background subtraction process and which may be subject to reddening
effects (see section 3), the best visual
match is with A0 to A3 dwarfs and class III giants.  The match is
illustrated in Fig.~\ref{fig3} in which selected comparison
stars from the stellar library of Jacoby \etal(1984) are presented. The
\pgccs~are {\it not} well--represented by class I or II giant star
spectra nor by spectra of stars of significantly cooler or hotter spectral
classification. For instance, there are no obvious HeI lines which rules
out the hottest stars. There
is a hint of G--band contribution to the blue wing of \Hgamma~which
suggests an A rather than late--B classification.  Types later
than about A3 seems to be ruled out because of their relatively
strong Ca K lines.  To quantify these impressions we have
measured equivalent widths in various A-- and B--type main sequence stars
from the compilations of both Jacoby \etal(1984) and Gunn \& Stryker
(1983).

It is worth noting here that there exist in the literature and in common
use a number of alternative bandpass definitions and methods of measuring
equivalent widths (EWs). The method employed in the Bruzual \& Charlot
(1993) models (hereafter BC93) gives systematically
larger values of Balmer EWs than the Brodie \& Hanes (1986) method (also
adopted by Zepf \etal1995).  While the Brodie \& Hanes bandpasses are
well--suited to measuring features in old globular clusters and early--type
galaxies, they are too narrow for measuring Balmer lines in young stellar
populations where these features are relatively strong and broad.  The
algorithm employed in the BC93 models for measuring the Balmer line
strengths is better suited to measuring the wide Balmer lines of young
stellar populations.  However, the choice of continuum level in the BC93
program is determined only by the flux values of the end points of
the continuum bandpasses and can, therefore, be greatly
affected by noise.  This algorithm is not appropriate for
low S/N spectra.  For this reason we have employed
Balmer line bandpass definitions that have been custom--defined for the
purpose of studying young stellar populations. These bandpasses (referred to
as LS indices after Linda Schroder who defined them) are
provided in Table 3. The algorithm we use to
determine the continuum level is based on the average flux value in the
continuum bandpasses and so is better suited to low S/N
spectra.  The LS bandpasses have been chosen to agree well
with the BC93 method when applied to the same 
high S/N spectra.  

\BC~(1993) models reflecting changes described in
Charlot, Worthey \& Bressan (1996), were circulated in 1995.
These newer models (hereafter BC95)
return EW values measured from SEDs with 10 \app~resolution.
This resolution is insufficient for accurate
measurement of narrow--bandpass indices and, in any case,
the very large widths of the Balmer lines
in young stellar populations make the use of such narrow--bandpass indices
inappropriate.  The more appropriate BC95 and LS indices
were measured on stellar spectra from the libraries of Jacoby \etal
(1984) and Gunn \& Stryker (1983) to produce a plot of
equivalent width as a function of spectral type (Fig.~\ref{fig4}).  The
Jacoby \etal spectra were first smoothed to match the 10 \app~resolution
of the Gunn \& Stryker spectra.  As demonstrated in
Fig.~\ref{fig4}, the agreement between LS and BC95 indices is excellent for
\Hbeta~and \Hdelta.  The small systematic offset in the values of
\Hgamma~is due to the fact that the BC95 algorithm only uses the red side
of the line when computing this equivalent width. This suppresses any
effect from of the G--band, which is present just blueward of \Hgamma.
Both BC95 and LS indices are relatively insensitive to spectral resolution.

\begin{figure}
\centerline{\psfig{figure=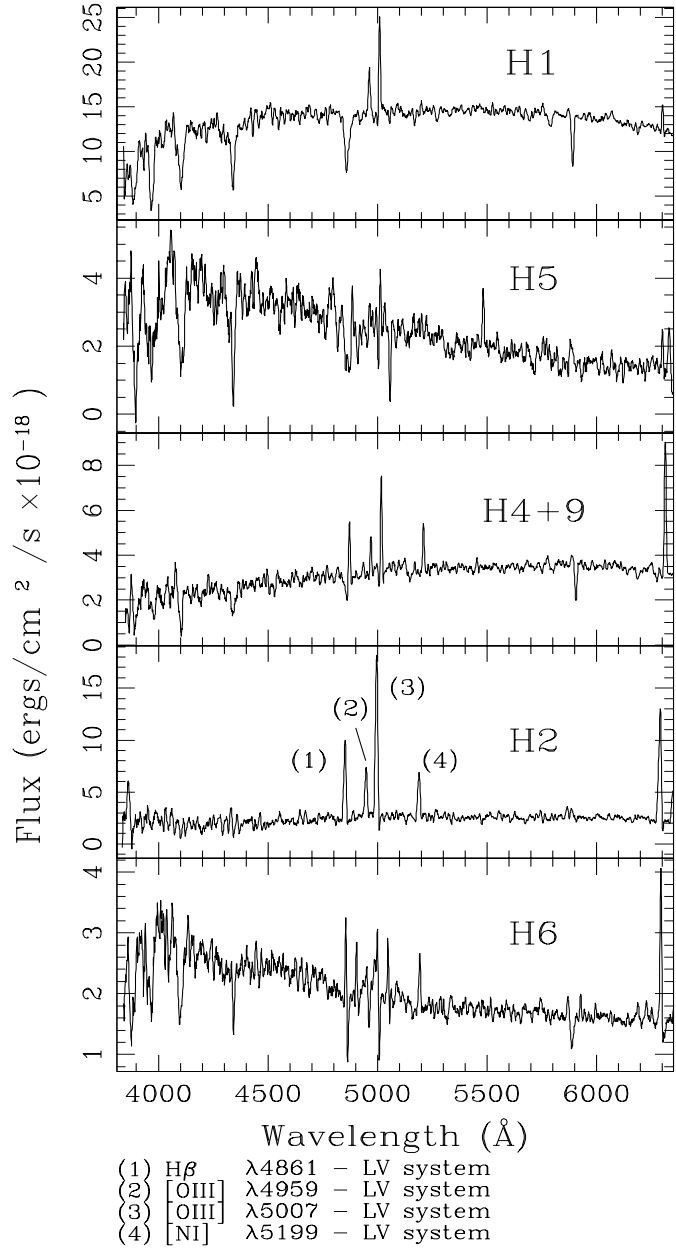,height=17.5cm,width=7.5cm,bbllx=72mm,bblly=50mm,bburx=144mm,bbury=237mm} }
\caption[fig2.eps]{\label{fig2} Keck I spectra of NGC 1275 \pgccs.
The data have been smoothed with a box size of seven pixels and the
resolution is 5.1 \AA.} The emission lines, which are residuals from
imperfect subtraction of the background galaxy/gas spectrum,
are marked to indicate the source of the line.  LV indicates
that the lines are residuals from the low--velocity system
(see text).  HV indicates that the lines are residuals
from the high--velocity infalling system (see text).
\end{figure}

\begin{figure}
\centerline{\psfig{figure=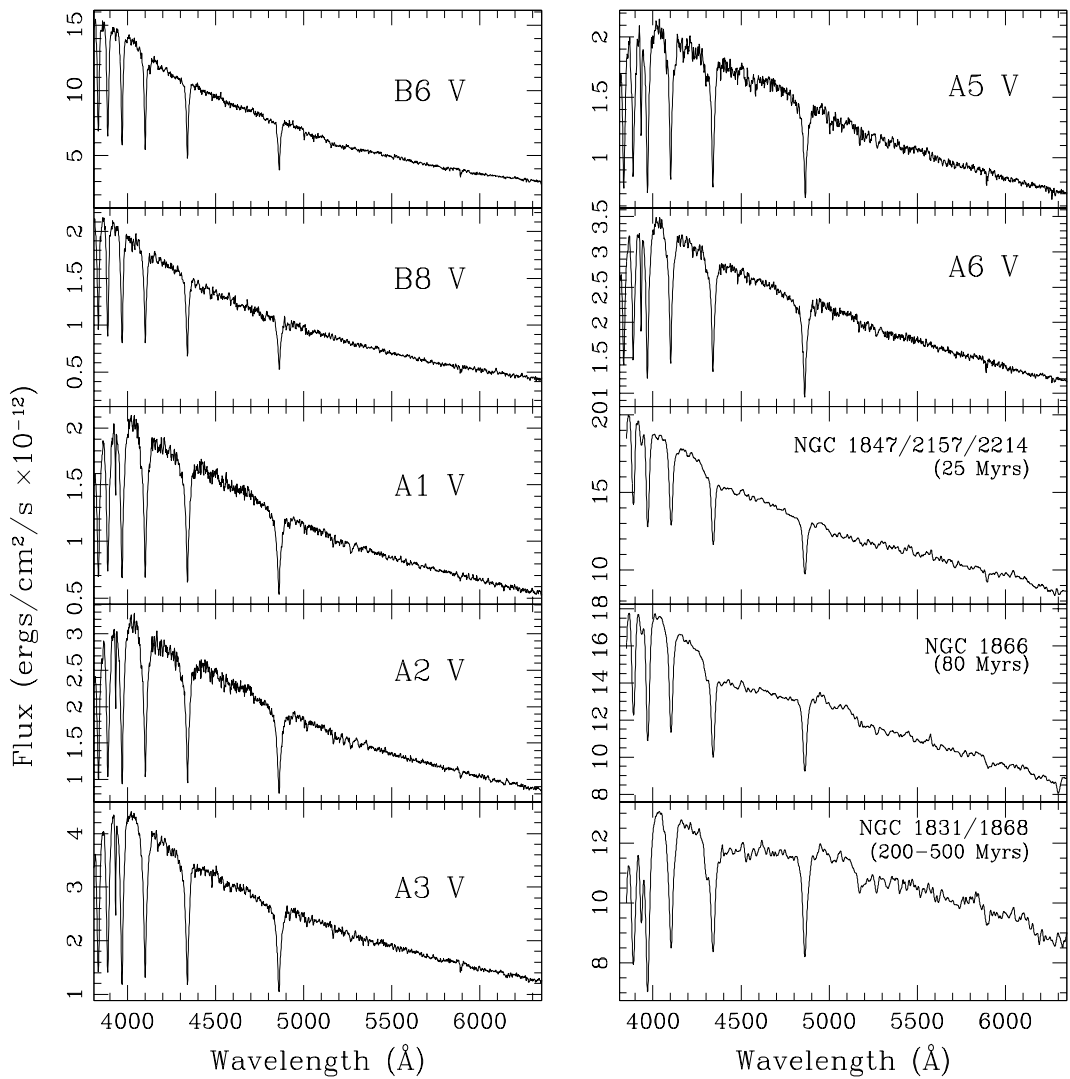,height=15.5cm,width=15cm,bbllx=20mm,bblly=64mm,bburx=199mm,bbury=242mm}}
\caption[fig3.eps]{\label{fig3} Spectra of solar metallicity main
sequence stars and integrated spectra of young star clusters for comparison
with H1.  We conclude that the optical light from
H1 is dominated by stars with a spectral type between
A0 and A3.  Of the integrated cluster spectra of Bica \& Alloin (1986),
H1 is most comparable to the composite LMC young globular cluster
NGC 1831/1868 with an of age of \agerangeM{200}{500} and [Fe/H] of -0.6,
but is not particularly well--represented by this spectrum.}
\end{figure}

\begin{figure}
\centerline{\psfig{figure=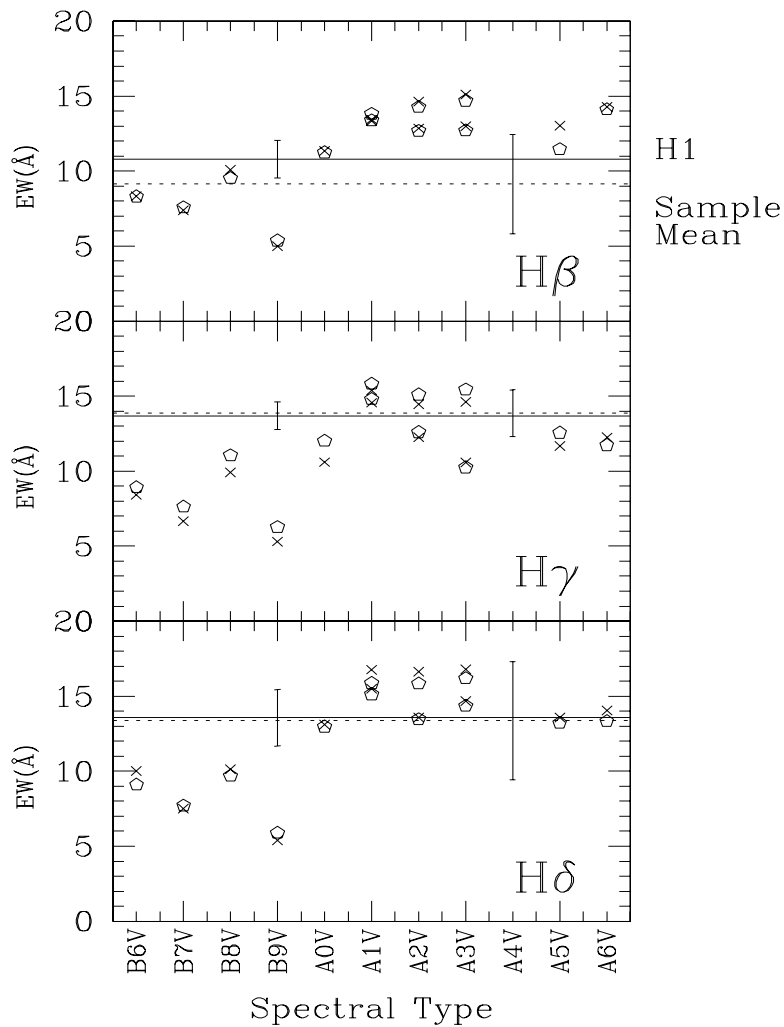,height=14.5cm,width=8cm,bbllx=61mm,bblly=85mm,bburx=140mm,bbury=242mm}}
\caption[fig4.eps]{\label{fig4} Balmer line equivalent widths of main
sequence stars vs.~spectral type, computed using two different algorithms.
The open pentagons represent the equivalent widths computed with the same
algorithm and bandpass definitions used in the BC95 models.  The crosses
represent equivalent widths computed using the method of Brodie \& Huchra
(1990) and the custom--defined LS bandpasses.  In both cases the spectra
were smoothed to 10 \app~to simulate the resolution of the BC95 model
spectra.  The two algorithms produce results that agree reasonably
well. Shown for comparison are horizontal lines indicating the Balmer line
equivalent widths (using the method of Brodie \& Huchra and the LS bandpass
definitions) of H1 (solid line) as well as the arithmetic mean for 4 of the
observed clusters (dotted line).  H2 was excluded from the mean, since the
Balmer line emission from the galaxy was clearly under--subtracted in that
spectrum.  The error bar for H1 represents the photon error.  The error bar
for the arithmetic mean represents the standard error on the mean, i.e.~the
dispersion divided by $\sqrt{N}$. }
\end{figure}

The horizontal lines in Fig.~\ref{fig4} represent the LS EW measurements of 
H1, for which we obtained the highest signal--to--noise data, and
the arithmetic mean of 4 clusters.  H2 was excluded from the mean since
the Balmer line emission from the galaxy was clearly
under--subtracted in that spectrum.  \Hbeta~is subject to the greatest 
background subtraction uncertainties of all the Balmer lines in our spectral
range because it is so strong in emission in the background spectrum. 
There is ample evidence in Fig.~\ref{fig2}, especially in the
lower signal--to--noise spectra, that \Hbeta~emission
from the background has been under--subtacted. The residual emission
serves to depress the mean EW for \Hbeta.  Note that, in general,
the background subtraction process was executed in such a way as to
result in either correct or under--subtraction of the background emission lines 
rather than their over--subtraction. 

Since the cluster spectra are integrated spectra, a perfect
match with a single spectral type is not to be
expected. Moreover, there is considerable scatter in Balmer EW measurements
for different stars within a given spectral type, although these
differences are generally comparable to our cluster measurement
errors.  However, the EWs confirm the visual impressions of the best
matching spectral type from Fig.~\ref{fig2} and Fig.~\ref{fig3}.
Taking account of the absence of strong Ca K, which rules out later
A--types, we conclude that the NGC 1275 \pgccs~are dominated
by early A--type stars with a best match at perhaps A0 or A1.
This conclusion is supported by the stellar evolutionary modeling described
in section 6.

Table 4 gives the LS EWs of \Hbeta, \Hgamma~and \Hdelta~for the NGC
1275 \pgccs.  For completeness, in Table 5 we provide
EWs for H1 derived using narrow--bandpass indices, for direct
comparison with the results of Zepf \etal(1995).
The definitions of the Mg{\it b} and Fe 5270
features are from Faber \etal(1985), and the Balmer line bandpasses are
those defined by Brodie \& Hanes (1986).

\section {Velocities}
Velocities were determined from each unfluxed spectrum using the IRAF
cross--correlation routine FXCOR.  Velocities quoted in Table 6 are
the average of the results of cross--correlation with 2 different,
high signal--to--noise, A--star templates with small (absolute values
of $<$\speed{7}) heliocentric velocities, obtained from the Jacoby
\etal (1984) library of stellar spectra plus a synthetic A--star
template made by combining spectra from a large number of different
A--stars. The cross--correlation was made over the region blueward of
\Hbeta~to avoid bias from possible under--subtraction of the
background emission lines. The Jacoby \etal spectra have a resolution
of 4.5 \AA~so could, in principle, provide a velocity precision of
$\sim$\speed{30}.  The results from the three templates agree quite
well, differences being a small ($\sim$10--30\%) fraction of the
velocity error.

We exploit the presence of Balmer absorption in the underlying galaxy
starlight to estimate the galaxy's ``local'' line--of--sight velocity
at the locations of the clusters.  At each cluster location the
cluster+galaxy+gas spectrum (from which the cluster spectra were
eventually extracted) was cross--correlated against the A--star
templates.  Since the light from the clusters amounts to, at most,
20\% of the total light producing these spectra, a cross--correlation
that is restricted to the Balmer lines should give a reasonable
estimate of the local bulk stellar velocity.  The local velocity of
the surrounding gas was estimated simply by measuring the line centers
of the best Gaussian fits to the various emission line profiles in the
cluster+galaxy+gas spectra.  The local galaxy and gas velocities
appear in Table 6. Excluding the composite object, H4+9, the mean
difference between the velocities of the clusters and the local
stellar velocity is \speedpm{47}{144}.  The mean difference
between the velocites of the clusters and the local gas velocity is
\speedpm{50}{101}.  There is no basis for preferentially associating
the clusters with either the gas or the stars.
None of the clusters is associated with the high velocity infalling material.

Zepf \etal(1995) found velocity differences for H1
``relative to the emission line gas'' of --120 and \speed{-140} 
at to two different position angles.  Our results
also show a comparable blue-shift of H1 with respect
to the [OIII] emission lines, but not with respect to
[OI] $\lambda 6300$ or \Hbeta.  It is not surprising that the gas velocities
yielded by these features are mildly discrepant.
Due to the wavelength dependence of dust scattering properties, 
measuring features at different wavelengths may be probing to different
depths along the line--of--sight. 
At the location of H2 and H5, the [OIII]
emission lines in the gas spectra appear to contain two velocity
components with the weaker component located blueward of the main component.
In these cases we measured the velocities of both components.
The resulting velocities are listed in Table 6.

We find a velocity dispersion of \speedpm{244}{99} for the clusters (excluding
the composite object, H4+9),
which, although very poorly constrained, seems to be
somewhat lower than that of globular cluster systems
in dominant central galaxies
(e.g. M87, Cohen \& Rhyzhov 1997; NGC 1399, Grillmair \etal1994,
Kissler-Patig \etal1998, Minniti \etal 1998).  
Note that the NGC 1275 
clusters are all very close to the galaxy
nucleus (within \about{2 kpc}).
The central stellar velocity dispersion of NGC 1275
has not been accurately measured but Lazareff \etal(1989) derived a value
of \speed{290} by modeling the mass distribution in the galaxy's
potential well. Recall that Lazareff \etal found a velocity dispersion
of \about{\speed{113}} for the CO gas in the center of NGC 1275, presumed
to be associated with the \Halpha~filaments and the clusters.
This value is marginally consistent with the cluster velocity dispersion
estimates, given the size of the errors.

The arithmetic mean velocity of the clusters, excluding H4+9, 
is \speedpm{5329}{121}, 
consistent with the galaxy velocity
of \speed{5264} measured by Strauss \etal(1992) as well as the local
galaxy velocity determined from the background spectrum.  The
error quoted is the standard error of the mean.

\section{Comparison with Star Clusters and Stellar Evolutionary Models}

\subsection{Star Clusters}

Bica \& Alloin (1986) present integrated spectra for star clusters
covering a wide range of ages and metallicities. Their plots of
index EW versus metallicity confirm the expected
lack of sensitivity to metallicity of line features in young cluster spectra.

Table 7 is a list of the LS Balmer line EWs for the spectra of
selected young clusters from the Bica \& Alloin data base, as tabulated by
Leitherer \etal(1996). The best match is obtained for clusters having
ages of 200--500 Myr, though the NGC 1275 clusters have
EWs which are larger than those for any cluster in
the Bica \& Alloin sample, irrespective of age or metallicity.
This suggests that the NGC 1275 clusters are
{\it not} well--represented by Magellanic Cloud young
or intermediate age clusters, nor by Galactic globular or open
clusters, a conclusion that is confirmed by visual inspection
of the spectra of the Bica \& Alloin objects (Fig.~\ref{fig3}).

\subsection{Bruzual \& Charlot Models}

\begin{figure}
\centerline{\psfig{figure=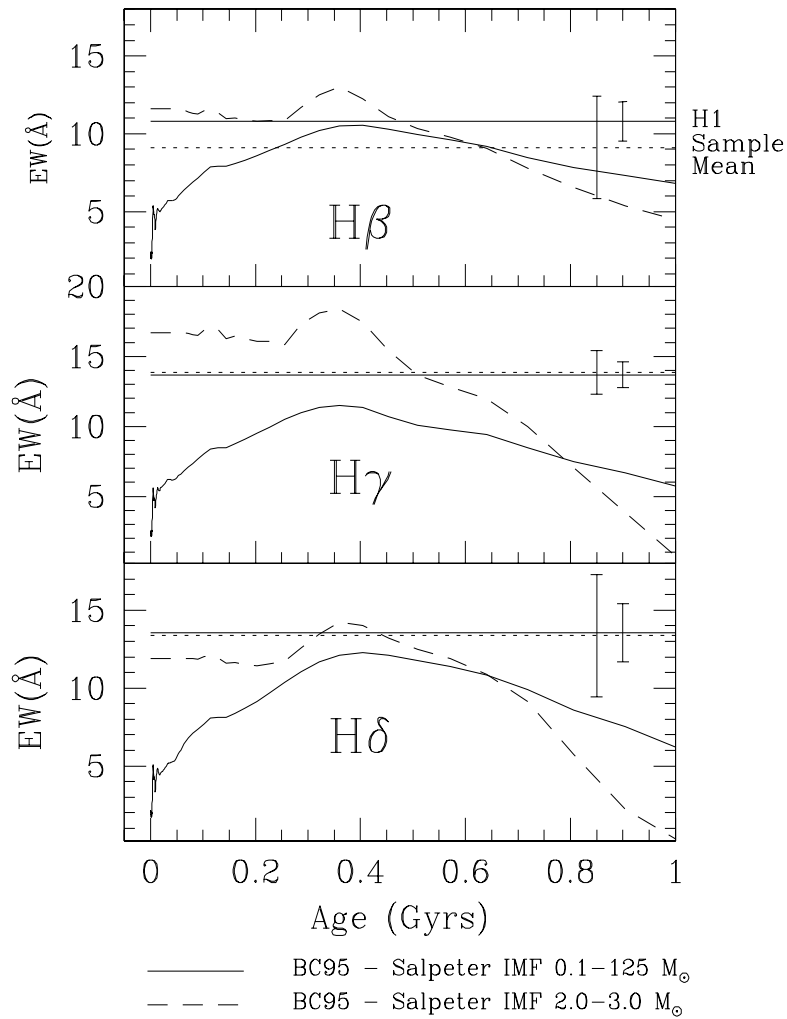,height=13.5cm,width=8.5cm,bbllx=56mm,bblly=89mm,bburx=142mm,bbury=219mm}}
\caption[fig5.eps]{\label{fig5} The time evolution of Balmer line equivalent
widths for populations
with different IMFs, produced by the BC95 models.  Also shown
are horizontal lines indicating the values of the Balmer
line equivalent widths of H1 (solid line)
and the arithmetic mean equivalent widths for 4 of the
observed clusters (dotted line).
H2 was excluded from the mean since the Balmer line emission
from the galaxy was clearly under--subtracted in that spectrum.
The error bar for H1 represents the photon error.  The error bar
for the arithmetic mean represents the standard error on the
mean, i.e.~the dispersion divided by $\sqrt{N}$.}
\end{figure}

BC95 stellar evolutionary models at solar metallicity
were used to generate plots of \Hbeta, \Hdelta~and \Hgamma~EWs as a
function of age. These are shown in Fig.~\ref{fig5}.
The horizontal lines in these plots represent the EWs
measured for the highest signal--to--noise cluster, H1, and 
the arithmetic mean of 4 clusters.  Again, H2 was excluded from
the mean because significant residual emission lines from the background galaxy
are present in its spectrum. Note that, as mentioned
in section 4, the mean EW of \Hbeta~is likely to be somewhat depressed
due to under--subtraction of the background \Hbeta~component. This
may be true for H1 but is especially relevant for the
lower signal--to--noise spectra.

Initially a Salpeter IMF (\Mrange{0.1}{125}~, X=1.35) was adopted and a
delta function burst of star formation at t=0 years was assumed.  
The relatively low S/N of our spectra (with the exception of H1)
prevents us from using the exact prescription of BC95 for a comparison of
the clusters' Balmer line strengths with the model predictions.  The
smoothing and rebinning to very low resolution (10 \app) combined with the
algorithm's ``end--point only'' determination of the continuum flux can
produce quite spurious results for low S/N spectra.  For this reason we
adopted the index measurement method outlined in Brodie and Huchra (1990)
using the LS bandpass definitions decribed in section 4.  Fig.~\ref{fig4}
demonstrates that, for high S/N spectra, this method and the BC95
algorithm produce very similar results. For low S/N spectra this method
allows a much more accurate determination of the continuum flux
and therefore provides the most robust comparison with the models.

In Fig.~\ref{fig5} the lines representing H1, the cluster with by far the
highest signal--to--noise spectrum, do not intersect the Salpeter IMF
curves.  It is possible to force an intersection for \Hbeta~
(known to be artifically depressed) and barely possible for
\Hdelta~(which is in the lowest
signal--to--noise portion of the spectrum) by pushing both the EWs
systematically lower towards the limit of their error bars.  The value for
\Hgamma, arguably the best--determined EW, is more than two standard
deviations away from the maximum of the Salpeter curve.  The sample mean
EWs for \Hgamma~and \Hdelta~also do not intersect the Salpeter IMF
curves. A shift towards the lower limit of its error bar does allow an
intersection for \Hdelta~but not for \Hgamma.
The \Hbeta~measurement is included for completeness but is not reliably
enough determined in these low signal--to--noise spectra to be a useful
constraint on the models.

\begin{figure}
\centerline{\psfig{figure=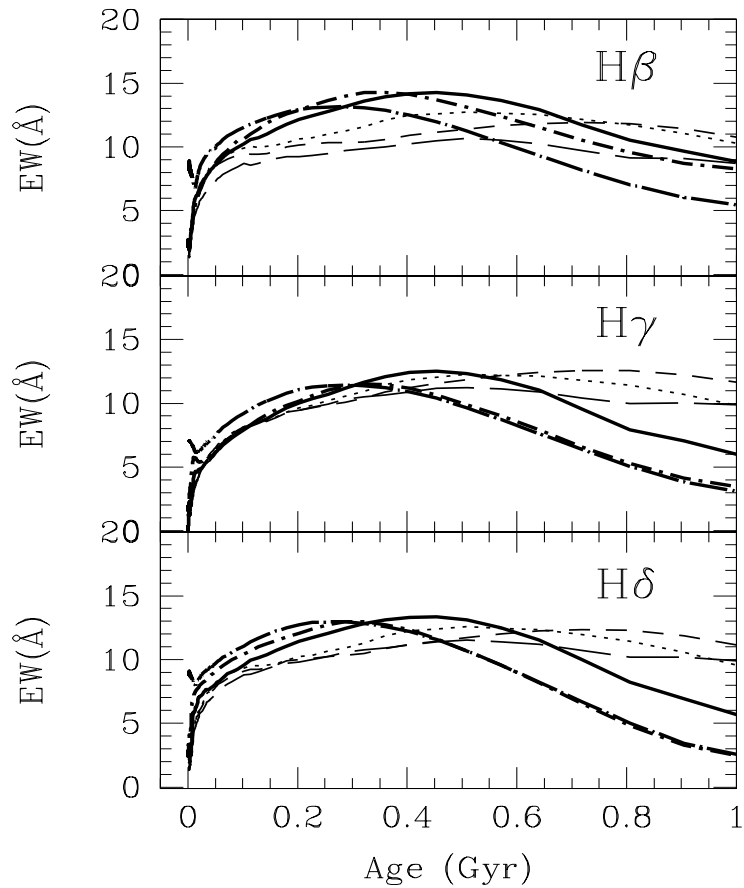,height=14.5cm,width=9.5cm,bbllx=50mm,bblly=92mm,bburx=142mm,bbury=243mm}}
\caption[fig6.eps]{ \label{fig6}  The time evolution of Balmer line equivalent
widths for a simple stellar populations of various metallicities, produced by
the preliminary BC97 models.  For ages
less than \about{0.5 Gyr}, \Hgamma~gives the tightest constraint on age, if the
metallicity is unknown. This
figure demonstrates the possible spread in Balmer line equivalent widths that
can be attributed to metallicity differences.
}
\end{figure}

The latest, albeit preliminary, \BC~(1997) models (hereafter BC97) 
allow metallicity to be varied in discrete steps
(corresponding to [Fe/H] = --1.7, --0.7, --0.4, 0.0, 0.4 and 0.7) while earlier
versions forced the assumption of solar metallicity for the system.  These BC97
models were kindly released to us in preliminary form in advance of
publication.  The theoretical input spectra used in BC97
have a substantially lower resolution (20 \app) than
the empirical spectra used in BC95.   This has caused (at
the time of this writing)
a systematic offset from the BC95 models, especially at young ages, in
the sense that the BC97 models return higher values of Balmer line EWs. 
Nevertheless,
the preliminary models can adequately demonstrate the range of EWs
that can be produced by varying metallicity.  Fig.~\ref{fig6} shows
the Balmer line EWs versus age for the range of metallicities provided in the
BC97 models.  
If the metallicity is unknown, \Hgamma~gives the tightest
constraint on age, for ages $<$500 Myrs, owing to its small
spread in EW for populations of different
metallicity.  For each Balmer line the maximum value of the EW occurs
at solar metallicity for ages less than 500 Myr.
Clearly, the solid curves in Fig.~\ref{fig5} (which represent 
solar metallicity) cannot be brought into agreement with the values
measured for the clusters simply by allowing metallicity to vary.
Except for the artifically depressed mean EW of \Hbeta, the Balmer line
EWs of the clusters are larger than {\it any} returned by
the BC models at any metallicity and agreement can only be achieved by
systematic downward shifts of all the Balmer EWs. 

It is worth noting here that the BC93 models and later variants suffered
from an error which transposed the column labels on the output of EW values
of \Hdelta~and \Hgamma. (Bruzual, 1997).  This has been corrected in our
plots.  Both Zepf \etal(1995) and Schweizer \& Seitzer (1993) compare their
measurements of Balmer line EWs in \pgccs~equivalent widths from these
models.  Both note that the (BC93, solar metallicity) models are unable to
produce EWs as strong as those they observed.

Having established that the large Balmer line EWs in the cluster spectra
are not reproduced by varying metallicity in the standard
models, we now explore the possibility that the IMF of these clusters
is different from the standard IMF and specifically that it is skewed to
high masses.  All versions of the \BC~models allow
for the adoption of Salpeter (1955) or Scalo (1986) IMFs
but neither of these is consistent with
the observed Balmer line strengths.  The input SEDs for the \BC~models,
as distributed, do not allow for variations in the slope of the IMF, nor can
the range of allowed masses be specified arbitrarily.  However, \BC~kindly
provided us with an input SED for their models, generated
originally for other purposes, which corresponds to a solar metallicity IMF
truncated to include only \Mrange{2}{3}~stars, i.e.~it has a low mass
cut--off at 2 \Msun~and a high mass cut--off at 3 \Msun.  This
essentially corresponds to a population of pure A stars.  We have
adopted this SED to simulate a flatter IMF. As is clear from
Fig.~\ref{fig5}, this model reproduces the strong Balmer line EWs of the
clusters and these are consistent with each
other in predicting an age of $\sim$\agerangeM{425}{500} for all the clusters
(with the mean \Hbeta~exception noted above). 

This age estimate is consistent with general arguments based on
main sequence lifetimes for early A--type stars. Poggianti \& Barbaro
(1997) confirm that the maximum Balmer line EWs occur in
A0V--type stars.  A0 star masses are estimated to be anywhere in the
range \Mrange{2.2}{3.5}~(e.g.~Allen 1974, Straizys \& Kuriliene 1981).
If higher mass stars are no longer present in significant numbers, a crude
age constraint can be determined from the duration of the H--burning phase
of \Mrange{2.2}{3.5}~stars. Such stars are predicted to 
stay on the main sequence for \about{\agerangeG{0.2}{0.7}},
depending on the isochrones adopted and the 
metallicities assumed (e.g. Schaller \etal1992)

\begin{figure}
\centerline{\psfig{figure=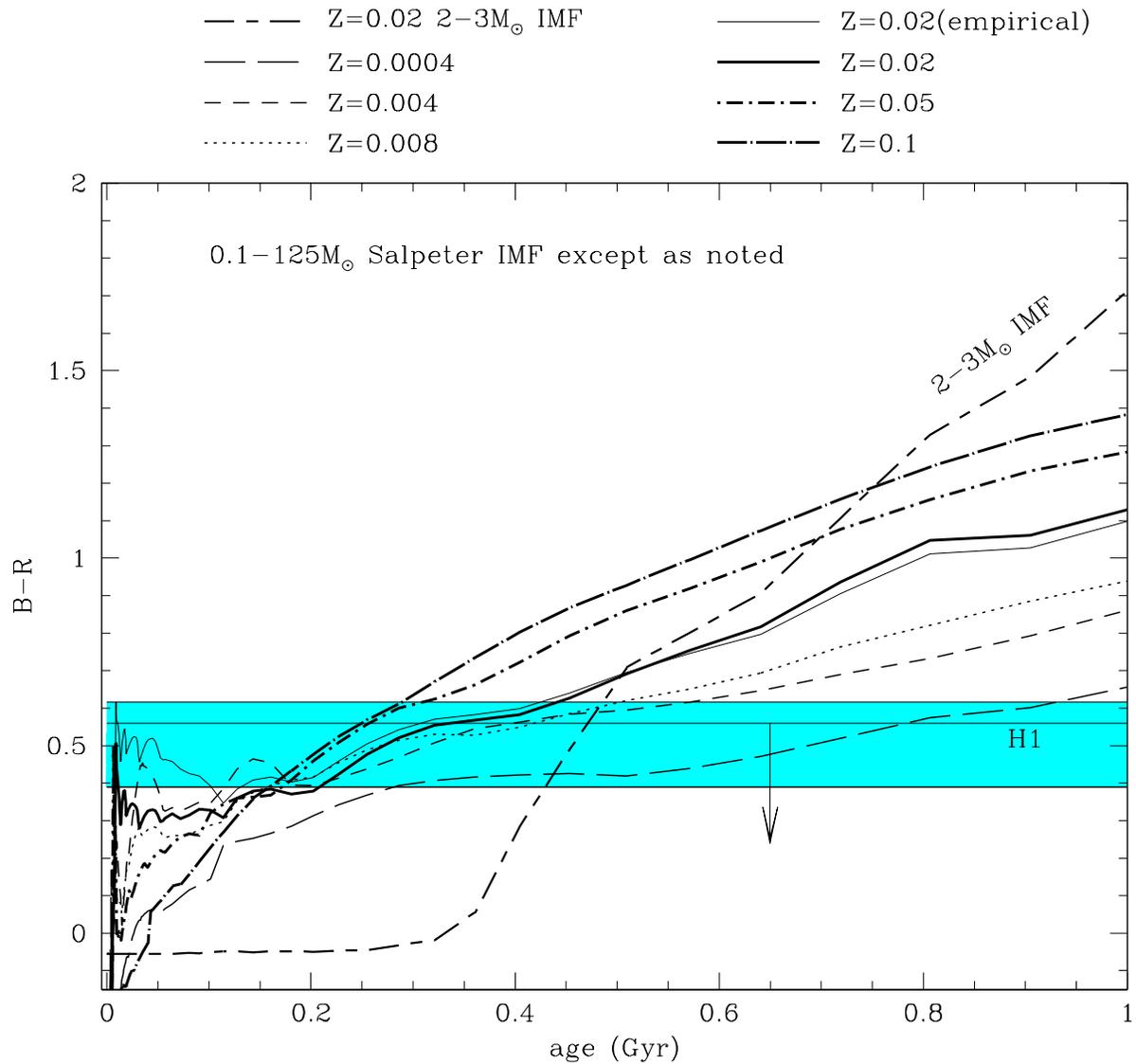,height=15.0cm,width=15.0cm,bbllx=22mm,bblly=58mm,bburx=199mm,bbury=238mm}}
\caption[fig7.eps]{\label{fig7} \color{B}{R} color vs.~Age produced by the BC97 preliminary models.
The color of H1, uncorrected for internal reddening,
is represented by a horizontal line. A downward--pointing arrow indicates the
effect of the estimated internal reddening, 
calculated from the HST images
as described in section 3. 
The possible range of the average color of the 5 clusters in the sample is
shown as a shaded region delimited by horizontal lines representing
colors corrected and uncorrected 
for internal reddening.}
\end{figure}

Preliminary BC97 models for \color{B}{R} as a function of age
are shown in Fig.~\ref{fig7}. Since
the \color{B}{R} colors measured from the empirical SEDs
of the older models and
the theoretical SEDs of the newest models appear to be consistent with
each other for this broad--band color, we have illustrated the dependence at
all available metallicities for the standard IMF (0.1--125$\,$\Msun) models.
Also shown is the \Mrange{2}{3}~model at solar metallicity.  
In Fig.~\ref{fig7} the color of H1, uncorrected for internal reddening,
is represented by a horizontal line. A downward-pointing arrow indicates the
effect of the estimated internal reddening, 
calculated from the HST images
as described in section 3. 
The average color of the 5 clusters in the sample, corrected and uncorrected
for internal reddening, is also shown.

From this figure we conclude that the most probable age of H1, deduced
from its color and the \Mrange{2}{3}~model is in the range 425--475 Myr, 
depending on the extent of the internal reddening. The corresponding
age range deduced from the average cluster colors is 400--475 Myr.
Note that the age derived from the color is consistent with
the age deduced from the Balmer line measurements.

If a standard (Salpeter) IMF is assumed, for all but the lowest metallicity
models, the age estimate drops to
400 Myr for the uncorrected color and less if {\it any} internal reddening
is present. The age estimate drops to an untenable (in view of the
spectral signature)  100 Myr if the
full internal reddening estimate is applied.
We have argued that the majority of the clusters are likely
to be suffering significant internal reddening. Moreover, 
significantly sub--solar
metallicities can be ruled out for ages as young as those derived from
the EW measurements.  

Taken as a whole, then, the cluster data appear to be more compatible with
a truncated and/or flattened IMF than with a Salpeter IMF. In particular,
age estimates from both EW and color measurements are mutually consistent
without the need to impose systematic shifts of the measured values or discount
internal reddening.

Although the agreement of our cluster data with the \Mrange{2}{3}~model
suggests that the IMF of the clusters
may be relatively flat, i.e.~skewed to high masses, it does not, by itself,
set any constraints on the nature of the IMF above 3 \Msun. After
\about{400 Myr}, only A--stars will be left in the clusters
regardless of the initial presence or absence of higher mass stars.
However, the characteristics of an original population of high mass
( $>$ 3 \Msun) stars, if any, can in principle be deduced
because such stars cannot be allowed to disrupt the clusters
prior to 400 Myr. In other words, the amount of mass lost from
the clusters caused by the evolution of massive stars must
be small enough that the clusters do not become unbound. Note too that, with
a 2 \Msun~lower mass cut--off, these clusters would fade away extremely
quickly (in \about{10$^9$ yr}), assuming they manage to avoid
disruption.  Future modeling efforts will explore variations in
the IMF slope and perhaps a variety of mass cut--offs in an attempt
to reproduce the observations even more closely.

\subsection{Fritze-v.~Alvensleben and Kurth Models}

Fritze-v.~Alvensleben \& Kurth (1997, hereafter FK) kindly compared our H1
spectrum with their models. Interestingly, the FK models can reproduce the
observed \Hbeta~and \Hgamma~EWs without the need to invoke a
non--standard IMF, although the agreement with \Hdelta~is less good.

Table 8 gives the Balmer line EWs and colors measured from model
SEDs of various ages.
FK models in the
range \agerangeM{200}{500} provide a reasonable fit to the observed EWs with
closest agreement for the 350 Myr model. For an age of 350 Myr, the mass
deduced for this cluster is \about{10$^7$ \Msun}. If the cluster is older the
mass estimate will increase, if younger it will decrease.  According to the
FK models, a 400 Myr old
cluster will fade by 3.4 magnitudes in 15 Gyr
while a 250 Myr old cluster
will fade by 3.8 magnitudes (O. Kurth, private
communication).  The current absolute B magnitude of H1 is --15.5, so in 15 Gyr
it will have M$_B$ \about{--11.9}, which is comparable to the brightest
globular clusters in M31, M87 and NGC 1399.  
H1 is by far the brightest of the \pgccs~but
it is clear on the basis of their luminosities,
and hence their masses, that none of these objects would be defined as
an open cluster. 

The {\it unreddened} color predicted by the 350 Myr FK model is
\color{B}{R}$\rm _0$=0.44. The B-R color of H1 is 0.56, uncorrected
for internal reddening, and is 0.24 if corrected for the
value of extinction estimated in the immediate vicinity of the cluster
for the dust lane in which H1 appears to lie.  
The model continuum is too red compared to the observed spectrum.
We argued previously that the precise continuum shapes in the observed spectra are
poorly defined because they are dependent on the background subtraction
process.  However, Fig.~\ref{fig2} and 
Fig.~\ref{fig3} indicate that the observed continuum shape is too red
for the inferred spectral type, rather than too blue. In addition,
no correction has been made for internal reddening. For these reasons, 
we infer that the true continuum shape of H1 is bluer than that appearing in
Fig.~\ref{fig2} and that appropriate corrections (for background subtraction
inaccuracies and internal reddening) would act to increase
the discrepancy between the model and the observation.
This discrepancy
is in the 
sense that the model predicts too much red light and is thus consistent
with the suggestion that the cluster IMF is flatter than a Salpeter IMF.

It is unclear why the BC95 and FK models give such different EW results.  Both
use the Padova evolutionary tracks (Bressan \etal1993) but there are subtle
differences between the two, such as the details of the template spectra
adopted.

\subsection{Metallicity}

Table 9 compares the strengths of Mg, Fe and Na features in the spectrum of
H1 with corresponding values from the 350 Myr FK models at various
metallicities. As expected for a young stellar population, the weak metal
lines are of little value in deducing the overall metallicity.  However,
despite the large errors, it is clear that Na is significantly overabundant
compared to its solar value.  The error on Mg is too large to judge its
strength clearly but it too may be overabundant.  We note that Na and Mg
overabundances have been seen in giant elliptical galaxies (Worthey
\etal1992) and in a small subset of globular clusters in NGC 1399
(Kissler-Patig \etal1998).  Worthey \etal suggest various mechanisms for
producing the Mg/Fe enhancements seen in the gE galaxies, one of which is a
flat IMF. 

\section {Discussion}

If we assume that the clusters in our
sample are representative of the system of \pgccs~as a whole, we
can draw some interesting conclusions from our results which pertain to our
understanding of the process of globular cluster formation
and set some constraints on the origins of globular clusters.

The mean velocity of the clusters is consistent with the velocity of
the main body of the galaxy and their velocity dispersion is \speedpm{244}{99}.
This dispersion is on the low side for globular clusters in giant
elliptical galaxies.  It has been argued that the clusters are associated with
the LV filaments, which in turn are associated with the CO gas. Both the
filaments and the CO gas are presumed to have been formed in the cooling
flow.  The cooling time for the CO gas is \about{10 Myr}, very similar
to the X--ray cooling time in these innermost regions (Mushotzky \etal 1981).
Since we can rule
out ages this young for the clusters, we can rule out their formation
in a {\it continuous} cooling flow and possibly, by association (if the
CO gas and the clusters are indeed spatially linked), the
formation of the LV filaments and the CO gas in the cooling flow.

\Halpha~images (Lynds 1970; Baade \& Minkowski 1954) indicate that the
extent of the filaments is 2--2.5\arcmin~for the LV system and
1--1.5\arcmin~for the HV system. Assuming a ``normal'' (King, de
Vaucouleurs, or Hubble--like) density profile for the \Halpha~material
would suggest an effective radius of perhaps 20--30\arcsec~and a core
radius of even less. The scale of the filaments might be similar to the
scale of the \pgccs.  The cooling flow radius is \about{5.5\arcmin} (Allen
\& Fabian 1997) which is a factor of 60 larger than the radius of our
sample and a factor of 20 larger than the entire sample of \pgccs.
Moreover, a cooling flow scenario requires star formation with a steep
IMF to account for the absence of observed high--mass stars in
conjunction with the deduced $>$ \Mdot{400} star formation rate. If
confirmed, the flat IMF (biased against low mass stars) deduced for
the clusters would be further evidence against their association with
the cooling flow. All this argues positively for a merger origin for
the filaments and the clusters.

The clusters show a remarkable degree of central concentration. From the
work of Holtzman \etal(1992) and Richer \etal(1993) we know that all the
currently--identified \pgccs~lie within \about{8 kpc} of the nucleus.
Richer \etal (1993) noted that the brighter clusters were more centrally
located than the fainter ones.  Those in our study, i.e.~the brighter
candidates, lie within \about{1.5 kpc} of the nucleus.  By contrast, the
old globular clusters lie within about 30 kpc of the nucleus (Kaisler
\etal1996), so the old and young clusters are very differently
distributed. This degree of central concentration appears to be greater
than that seen for metal--rich globular clusters in any ``normal''
elliptical galaxy.

It is not clear why formation in a merger would preferentially form
clusters only in the central few kpc if star formation is triggered in
chaotically distributed shocks. However, it is to be expected that gas from
merging galaxies will be channeled rapidly to the center 
(Barnes \& Hernquist 1996 and references therein) where
it can form stars very soon after the
onset of the merger event. It is unlikely that
widespread shocks from merging galaxies could 
have produced these clusters both on the grounds of their central
concentration and because their velocity dispersion may be relatively 
low. If the clusters
were formed further out in the galaxy and later fell into the center, a
high velocity dispersion would be expected. In fact,
it might be difficult to find a mechanism for moving the clusters
nearer to the galaxy center since dynamical friction is not
likely to be effective at these masses, although they might
conceivably have been associated
with an infalling galaxy. In any case, we can conclude that they
actually formed very close to the galaxy center.  A cooling flow or central
accumulation of merger gas is then more appealing in this regard because
the gravitational potential definitely steepens rapidly near the galaxy
center and objects formed in this potential would acquire the mean velocity
of the galaxy.  It will be of great interest to obtain similar
high--quality spectra of \pgccs~in other interacting galaxies,
especially ones without cooling flows.

If mergers are to contribute significantly to high S$_N$ globular cluster
systems (although see Forbes, Brodie \& Grillmair 1997 for arguments
against that idea) and mergers produce very centrally concentrated globular
clusters, some mechanism has to be invoked to redistribute the clusters to
resemble the observed spatial distributions in normal cluster systems.  For
example, in M87 (Grillmair \etal 1986 and Lauer \& Kormendy, 1986) and in
NGC 1399 (Kissler-Patig \etal 1997) it has been established that the
globular cluster systems are more diffuse than the galaxy light. A similar
argument can be made for centrally concentrated objects formed from a
cooling flow.

The crossing time for objects with \about{\speed{200}} velocity dispersion
with a radius of 8 kpc (the radial extent of whole \pgcc~sample) is
\about{40 Myr}, or a factor of 10 less than their age. Their survival in
the galaxy center must indicate that they are in virial equilibrium,
unless the velocity dispersion measurement is unrepresentative because of
the very small sample of velocities. 

We have found evidence that these clusters have an IMF that is flat
(skewed to high masses or biassed against low masses) 
compared to a standard IMF.  The low-mass stellar
component may be weak or possibly absent in these objects and, in order to
avoid disruption long enough to survive to $\sim$400 Myr, there are
restrictions on the high mass component.

Interestingly, recent work by Padoan \etal(1997) provides a natural framework
for such a result if star formation is driven by random supersonic
flows. They predict that starburst regions should have flatter IMFs with a
more massive low--mass cut-off because of their higher mean temperatures.
Based on core--radius measurements, Elson, Freeman \& Lauer (1989) report
evidence for a flatter IMF in their study of the core radii of young
clusters in the Large Magellanic Cloud. However, as already noted, the
spectra of LMC clusters are not similar to those of the NGC 1275 young
clusters.

\section{Summary and Conclusions}

The clusters we have studied are not HII regions and their
integrated spectral properties do not
resemble those of 
young or intermediate age Magellanic Cloud clusters or Milky Way
open clusters.

The clusters' Balmer line strengths appear to be too strong to be consistent
with any of the standard \BC~evolutionary models at any 
age or metallicity. If the
\BC~models are adopted, an IMF which is skewed
to high masses provides
a more internally consistent fit to the spectral and photometric data. 
A truncated IMF with a mass range of \Mrange{2}{3}, adopted to mimic
a flatter IMF, reproduces
the observed EWs at \agerangeM{425}{500} and the cluster colors
predict ages of \about{\agerangeM{400}{475}} for this mass range.  

The FK models are better able to reproduce the observed Balmer line
strengths for a solar metallicity cluster with an age of 350 Myr and a
{\it standard} (Salpeter) IMF, although the model continuum is too red
compared to the observed spectrum.  The continuum discrepancy between
the model and the observation is in the sense that the model predicts
too much red light and thus is also consistent with the suggestion
that the cluster IMF is relatively flat.

Key properties of the clusters, in addition to their possibly flat IMF, are
their extremely centrally concentrated spatial distribution, their 
velocity dispersion of \speedpm{244}{99} and their age of \about{450 Myr}.

Formation in a {\it
continuous} cooling flow appears to be ruled out since the age of the clusters
is much larger than the cooling time of the molecular gas (assuming that
the molecular gas is associated with the cooling flow), the spatial scale of
the clusters is much smaller than the cooling flow radius, and the deduced star
formation rate in the cooling flow favors a steep rather than a flat IMF.

A merger would have to produce clusters only in the central few kpc,
presumably from gas from the merging galaxies which is channeled rapidly
to the center.  Widespread shocks from merging galaxies cannot have
produced these clusters.  If they were formed further out in the galaxy
and later fell into the center (by some unspecified mechanism), 
a high velocity dispersion would be
expected.

The crossing time of the clusters is only \about{40 Myr}. Their
survival to \about{400 Myr} indicates that they must be in virial equilibrium
if the observed velocity dispersion is representative of the whole
young cluster population.

Regardless of their origin, these objects are not distributed like
old globular clusters in central cD galaxies where the globular
clusters are significantly more diffuse than the galaxy light.

The clusters appear to have formed in a discrete event some 450 Myr
ago. This event might have been induced by a merger which provided the fuel
for a short--lived gas infall episode.  With a small or absent low mass
stellar component, these objects would fade away very rapidly (in
\about{$10^9$ yr}), assuming they can avoid disruption in the later stages
of stellar evolution.  More data and further modeling, especially to
determine the slope and/or cut--off parameters of the IMF, are required to
ascertain whether or not they will eventually evolve into objects we would
regard as {\it bona fide} globular clusters.\\

\acknowledgments 

We thank Harvey Richer for providing accurate coordinates for the
\pgccs~and for useful discussions and we are
indebted to Oliver Kurth for comparing our spectra with the
Fritze-v.~Alvensleben \& Kurth models and for many
insightful comments.  We have benefited from useful
discussions with Ann Zabludoff, Ken Freeman, Dennis
Zaritsky, Peter Bodenheimer, Bill Matthews, Uta Fritze-v.~Alvensleben
and especially Doug Lin. We are very grateful to Jon Holtzman for
helping us get the colors right and for valuable suggestions. We
appreciate the many helpful comments of the anonymous referee. 
We would like to thank the staff at the Keck Observatories, as well as the
entire team of people, led by J.B. Oke and J.G. Cohen, responsible for the
Low Resolution Imaging Spectrograph, for making the observations possible.
 
This research was funded in part by HST grant
GO.05990.01-94A, the Smithsonian Institution and by faculty research funds
from the University of California at Santa Cruz.\\

\leftline{\bf REFERENCES}\vspace{2mm}

\noindent
Arnaud, K.A., \etal 1994, ApJ, 436, L67\\
Ashman, K.M., \& Zepf, S.E.  1992, ApJ, 384, 50\\
Allen, C.W. 1974, Astrophysical Quantities, 3rd Ed., (The Athalone Press, 
London), p. 208 \\
Allen, S.W., \& Fabian, A.C. 1997, MNRAS, 286, 583 \\
Allen, S.W., Fabian, A.C., Johnstone, R.M., Nulsen, P.E.J., \& Edge, A.C. 1992, MNRAS, 254, 51\\
Baade, W., \& Minkowsky, R. 1954, ApJ 119, 215\\
Barnes, J.E. \& Hernquist, L. 1996, ApJ 471, 115 \\
Bica, E., \& Alloin, D. 1986, A\&A, 162, 21\\
Blakeslee, J. 1996, PhD thesis, M.I.T.\\
Bressan, A., Fagotto, F., Bertelli, G., \& Chiosi, C. 1993, A\&AS, 100, 647\\
Brodie, J.P., \& Hanes, D.A. 1986, ApJ, 300, 258\\
Brodie, J.P., \& Huchra, J.P. 1990, ApJ, 362, 503\\
Burstein, D., \& Heiles, C. 1984, ApJS, 54, 33 \\
Burstein, D., Faber, S., Gaskell, M., \& Krumm, N. 1984, ApJ, 287, 586\\
Bruzual, G., \& Charlot, S. 1993, ApJ, 405, 538 \\
Bruzual, G., \& Charlot, S. 1997, in preparation\\
Bruzual, G., 1997, private communication \\
Charlot, S., Worthey, G., \& Bressan, A. 1996, ApJ, 457, 625\\
Cohen, J.G., \& Rhyzhov, A.S. 1997, AJ 486, 230\\
Elson, R.A.W., Freeman, K.C. \& Lauer, T.R. 1989 ApJ 347, L69\\
Faber, S., Friel, E.D., Burstein, D., \& Gaskell, C.M. 1985, ApJS, 57, 711\\
Faber, S. 1993,  The Globular Cluster--Galaxy Connection, 
ASP conf. series Vol. 48, eds. G. Smith, \& J. Brodie, p.601\\
Fabian, A.C., Nulsen, P.E.J., \& Canizares, C.R. 1984, Nature, 310, 733\\
Fabian, A.C., Hu, E.M., Cowie, L.L., \& Grindlay, J. 1981, ApJ, 336, 734\\
Ferruit, P., \& P\'{e}contal, E. 1994, A\&A, 288, 65\\
Forbes, D., Brodie, J.P., \& Huchra, J.P. 1996, AJ, 112, 2448\\
Forbes, D., Brodie, J.P., \& Grillmair, C. 1997, AJ, 113, 1652\\
Fritze-v.~Alvensleben, U. \& Burkert, A. 1995, A\&A, 300, 58\\
Fritze-v.~Alvensleben, U. \& Kurth, O. 1997, in preparation\\
Grillmair, C., Pritchet, C., \& van den Bergh, S. 1986, AJ, 91, 1328 \\
Grillmair, C.,  \etal 1994, ApJ, 422, L9\\
Gunn J.E., \& Stryker, L.L. 1983, ApJS, 52, 121\\
Harris, W.E. 1991, ARA\&A, 29, 543\\
Holtzman, J.A., \etal 1992, AJ, 103, 691\\
Holtzman, J.A., \etal 1995, PASP, 107, 1066\\
Holtzman, J.A., \etal 1996, AJ, 112, 416\\
Holtzman, J.A., \etal 1998, in preparation\\
Johnson, H.L. 1966, ARAA, 4, 193\\
Jacoby, G.H., Hunter, D.A., \& Christian, C.A. 1984, ApJS, 56, 278\\
Kaisler, D., Harris, W.E., Crabtree, D.R. \& Richer, H.B. 1996, AJ, 111, 2224\\
Kissler-Patig, M., Kohle, S., Hilker, M., Richtler, T., Infante, L., Quintana, H. 1997, A\&A, 319, 470\\
Kissler-Patig, M., Brodie, J.P., Schroder, L.L.,Forbes, D.A., Grillmair, C.G.,
\& Huchra, J.P. 1998, AJ 115, 105\\
Lauer, T., \& Kormendy, J. 1986, ApJ, 303, L1\\
Lazareff, B., Castets, A., Kim, D.W., \& Jura, M. 1989, ApJ, 336, L13\\
Leitherer, C. \etal 1996, PASP, 108, 996\\
Lynds, C.R. 1970, ApJ, 159, L151\\
Miller, B.W., Whitmore, B.C., Schweizer, F., \& Fall, S.M. 1997, AJ in press\\
Minkowski, R. 1957, Radio Astronomy, IAU Symposium No. 4, ed. H.C.
van de Hulst (Cambridge University Press, Cambridge), Vol. 107\\ 
Minniti, D., Kissler-Patig, M., Goudfrooij P., \& Meylan, G. 1998, AJ,
115, 121\\
Mushotzky, R. 1992, Clusters and Superclusters of Galaxies, ed. A.C.
Fabian (Kluwer, Dordrecht), p. 91\\
Mushotzky, R., Holt, S.S., Boldt, E.A., Selemitsos, P.J., \& Smith, B.W. 1981,
ApJ 244, L47\\
N{\o}rgaard-Nielsen, H.U., Goudfrooij, P., Jorgensen, H.E., \& Hansen, L. 1993, AA, 279, 61\\
Oke, J.B., et al., 1995, PASP 107, 375\\
Padoan, P., Nordlund, A., \& Jones, B.J.T. 1997, MNRAS 288, 145\\
Poggianti, B.M., \& Barbaro, G. 1997, A\&A, in press\\
Richer, H.B., Crabtree, D.R., Fabian, A.C., \& Lin, D.N.C. 1993, AJ, 105, 877\\
Rieke, G.H., \& Lebofsky, M.J. 1985, ApJ, 288, 618\\
Salpeter, E.E. 1955 ApJ, 121, 61\\
Scalo, J.M. 1986, Fundamentals of Cosmic Physics, 11, 3 \\
Schaller, G., Schaerer, D., Meynet, G., \& Maeder, A. 1992, A\&AS, 96, 269\\
Schweizer, F., \& Seitzer, P. 1993, ApJ, 417, L29\\
Schweizer, F. 1987, Nearly normal galaxies: From the Planck time
to the Present, Proceedings of the Eighth Santa Cruz Summer Workshop
in Astronomy and Astrophysics, ed. S. Faber (Springer-Verlag, New York), 
p.18 \\
Smith, E.P. \etal 1992, ApJ, 395, L49\\
Straizys, V., \& Kuriliene, G. 1981, Ap\&SS, 80, 353 \\
Strauss, M., Huchra, J.P., Davis, M., Yahil, A., Fisher, K., \& Tonry, J. 1992, ApJS, 83, 29\\
Straziys, V. \& Kuriliene, G. 1981, Ap\&SS, 80, 353\\
Whitmore, B., Schweizer, F., Leitherer, C., Borne, K., \& Robert, C. 1993, AJ, 106, 1354\\
Whitmore, B.C., \& Schweizer, F. 1995, AJ, 109, 960\\
West, M.J., Cote, P., Jones, C., Forman, W., \& Marzke, R.O. 1995 
ApJ 453, L77\\
Worthey, G., Faber, S.M., \& Gonzales, J.J. 1992, ApJ 398, 69 \\ 
Zepf, S.E., Carter, D., Sharples, R.M., \& Ashman, K.M. 1995, ApJ, 445, L19\\
Zabludoff, A., Huchra, J.P., \& Geller, M. 1990, ApJS, 74, 1\\

\clearpage

\begin{deluxetable}{lrrr}
\tablewidth{0pc}
\tablenum{1}
\tablecaption{Observational Data}
\tablehead{
\colhead{Cluster}  & \colhead{Slit}  & \colhead{Integration} & \colhead{Dates}   \nl
\colhead{ID}   & \colhead{PA}    & \colhead{Time}        & \colhead{Observed}\nl 
\colhead{(1)}      & \colhead{(2)}   & \colhead{(3)}         & \colhead{(4)}     }
\startdata
H1      &  153.0  &   900   &  Dec 01 \nl	
\nodata &   46.7  &  6300   &  Nov 30 \nl
H2      &  113.5  &  3200   &  Nov 30 \nl
H3      &   46.7  &  6300   &  Nov 30 \nl
H4+H9   &  113.5  &  3200   &  Nov 30 \nl
H5      &  113.5  &  3200   &  Nov 30 \nl
\nodata &  158.1  &  1800   &  Dec 01 \nl
\nodata &  140.0  &   900   &  Dec 01 \nl
H6      &   46.7  &  6300   &  Nov 30 \nl
\nodata &  158.1  &  1800   &  Dec 01 \nl
H12     &   77.4  &  4500   &  Dec 01 \nl
H17     &   77.4  &  4500   &  Dec 01 \nl
\enddata
\tablenotetext{(1)}{Cluster identification number,}
\tablenotetext{   }{   as taken from Holtzman \etal (1992)}
\tablenotetext{(2)}{Degrees from counter-clockwise from N-S}
\tablenotetext{(3)}{Exposure time in seconds}
\tablenotetext{(4)}{Date of Observation, 1994}
\end{deluxetable}

\begin{deluxetable}{lrccrrrcrlcrrrr}
\tablewidth{0pc}
\tablenum{2}
\tablecaption{Magnitudes$^{*}$ and Colors}
\tablehead{  & \multicolumn{2}{c}{This}                    &&
               \multicolumn{1}{c}{F \& P} 		   &&		  
	       \multicolumn{3}{c}{Richer}                  &&
               \multicolumn{2}{c}{Holtzman}                &&
               \multicolumn{2}{c}{N\o rgaard}              \\
	     & \multicolumn{2}{c}{Paper}                   &&
               (1994)                                      &&
  	       \multicolumn{3}{c}{\etal (1993)}            &&
               \multicolumn{2}{c}{\etal (1992)}            &&
               \multicolumn{2}{c}{\etal (1993)}            \\
             & \multicolumn{2}{c}{(1)}                     &&
               \multicolumn{1}{c}{(2)}                     &&
               \multicolumn{3}{c}{(3)}                     &&
               \multicolumn{2}{c}{(4)}                     &&
	       \multicolumn{2}{c}{(5)}                     \\
\cline{2-3} \cline{5-5} \cline{7-9} \cline{11-12} \cline{14-15}\\
\colhead{Cluster ID} & \colhead{B} & \colhead{R} &&
\colhead{R}  &&
\colhead{B} & \colhead{V} & \colhead{I} &&
\colhead{V} & \colhead{R} &&
\colhead{V} & \colhead{I} \\ }
\startdata
H1  & 19.00 & 18.44               &&
      18.50                       &&
      18.98 & 18.65 & 18.38       &&
      18.41 & 18.22               &&
      18.36 & 17.73              \nl
$\sigma_{H1}$
    &  0.15 &  ~$\,$0.15		  &&
       0.16  			  &&
       0.01 &  ~$\,$0.03 &  0.02       &&
       ~$\,$0.02 &  0.02               &&
       0.01 &  0.00              \nl
H2  & 20.18 & 19.53               &&
      19.54                       &&
      20.11 & 19.89 & 19.53       &&
      19.51 & 19.38               &&
      19.95 & 19.43              \nl
$\sigma_{H2}$
    &  0.15 &  ~$\,$0.15		  &&
       0.16  			  &&
       0.07 &  ~$\,$0.09 &  0.09       &&
       ~$\,$0.08 &  0.14               &&
       0.02 &  0.02              \nl
H3  & 20.05 & 19.35               &&
      19.62                       &&
      20.34 & 19.66 & 19.48       &&
      19.41 & 19.23               &&
      \nodata &\nodata           \nl
$\sigma_{H3}$
    &  0.15 & ~$\,$0.15		  &&
       0.35  			  &&
       0.11 & ~$\,$0.04 &  0.04        &&
       ~$\,$0.08 & 0.05                &&
      \nodata &\nodata           \nl
H5  & 20.57 & 20.15               &&
     \nodata                      &&
      20.71 & 20.36 & 20.05       &&
      20.11 & 20.01               &&
      20.96 & 20.56              \nl
$\sigma_{H5}$
    &  0.15 &  ~$\,$0.15		  &&
      \nodata 			  &&
       0.05 & ~$\,$0.06 &  0.03        &&
       ~$\,$0.04 & 0.06                &&
       0.03 & 0.05               \nl
H6  & 20.76 & 20.04               &&
     \nodata                      &&
      20.63 & 20.46 & 19.91       &&
      20.21 & 20.02               &&
      20.12 & 19.45              \nl
$\sigma_{H6}$
    &  0.15 &  ~$\,$0.15		  &&
       \nodata			  &&
       0.19 &  ~$\,$0.08 &  0.03       &&
       ~$\,$0.06 &  0.08               &&
       0.02 &  0.00              \nl
H4+H9$^{**}$
     & 19.72 & 19.07              &&
      20.04                       &&
   $($19.91 & 20.52 & 20.70$)$    &&
   $($19.48 & 19.30$)$            &&
   $($20.21 & 19.74$)$           \nl
$\sigma_{H4+H9}$
    &  0.15 & ~$\,$0.15		  &&
       0.47   	 	          &&
    $($0.06 & ~$\,$0.18 & 0.22$)$      &&
    $($0.08 & 0.11$)$             &&
     $($0.02 & 0.02$)$            \nl
& & & & & & & & & & & & & &      \nl
     & B--R & E$_{(B-R)}$ &&	  &&
       B--V & V--I &	          &&
       V--R &                     &&
       V--I  	                   &
	                         \nl
H1   & 0.56 & 0.32 &&             &&
       0.33 & 0.27 &              &&
       0.19 &                     &&
       0.63                      \nl
$\sigma_{H1}$
     & 0.21 &      &&             &&
       0.03 & 0.04 &              &&
       0.03 &                     &&
       0.01                      \nl
H2   & 0.65 & 0.18 &&             &&
       0.22 & 0.36 &              &&
       0.13 &                     &&
       0.39                      \nl
$\sigma_{H2}$
     & 0.21 &      &&             &&
       0.11 & 0.13 &              &&
       0.16 &                     &&
       0.03                      \nl
H3   & 0.70 & 0.23 &&             &&
       0.68 & 0.18 &              &&
       0.18 &                     &&
       \nodata                   \nl
$\sigma_{H3}$
     & 0.21 &      &&             &&
       0.12 & 0.06 &              &&
       0.09 &                     &&
       \nodata                   \nl
H5   & 0.42 & 0.03 &&             &&
       0.35 & 0.31 &              &&
       0.10 &                     &&
       0.40    	 	         \nl
$\sigma_{H5}$
     & 0.21 &      &&             &&
       0.08 & 0.07 &              &&
       0.07 &                     &&
       0.06                      \nl
H6   & 0.72 & 0.25 &&             &&
       0.17 & 0.55 &              &&
       0.19 &                     &&
       0.67                      \nl
$\sigma_{H6}$
     & 0.21 &      &&             &&
       0.21 & 0.09 &              &&
       0.10 &                     &&
       0.02                      \nl
H4+H9 & 0.65 & 0.33 &&             &&
       (-0.60 & -0.18) &             &&
       (0.18)  &                     &&
       (0.47)                       \nl
$\sigma_{H4+H9}$
     & 0.21 &       &&             &&
       (0.19 & 0.28) &              &&
       (0.14) &                     &&
       (0.03)                      \nl
\enddata
\tablenotetext{NOTES}
\tablenotetext{}
\tablenotetext{*}{Magnitudes have been corrected for foreground Galactic extinction (i.e. A$_B=0.71$, A$_V=0.54$, A$_R=0.40$ and A$_I=0.26$).  The Holtzman \etal values have been corrected for the acknowledged (Richer \etal 1993) factor of 2 error, i.e. 0
.75 mags. \nl}
\tablenotetext{**}{The objects H4 and H9 are unresolved from the ground.
Ferruit \& P\'{e}contal (1994) note this and give
a magnitude for H4+H9.  Richer \etal(1993),
however, list magnitudes only for H9, whereas N\o rgaard--Nielsen \etal(1993)
list magnitudes only for H4.  We have placed those magnitudes
on the line for H4+H9, since each is bound to include both objects.
The entries for H4+H9 in the Holtzman \etal(1992) columns were derived
by adding the objects' individual magnitudes. In this paper we have 
used an aperture encompassing both objects.  
\nl}
\tablenotetext{}{$^1$WFPC2, $^2$CFHT, $^3$CFHT, $^4$WFPC1, $^5$NOT}
\end{deluxetable}

\begin{deluxetable}{lccc}
\tablewidth{0pc}
\tablenum{3}
\tablecaption
  {Custom-defined LS Balmer Line Bandpasses}
\tablehead{
  \colhead{Feature} &
  \colhead{Blue}    & &
  \colhead{Red}      \\
  \colhead{Name}    &
  \colhead{Continuum} &
  \colhead{Feature} &
  \colhead{Continuum} }
\startdata
  H$\delta_{LS}$ & 4025.0--4060.0 & 4070.0--4135.0 & 4145.0--4180.0 \nl
  H$\gamma_{LS}$ & 4225.0--4275.0 & 4320.0--4370.0 & 4375.0--4425.0 \nl
  H$\beta_{LS}$  & 4800.0--4830.0 & 4830.0--4900.0 & 4905.0--4930.0 \nl
\enddata
\end{deluxetable}

\begin{deluxetable}{lrrrr}
\tablewidth{0pc}
\tablenum{4}
\tablecaption{Balmer Line Equivalent Widths}
\tablehead{
\colhead{Cluster}  &
\colhead{Position} & & & \\
\colhead{ID}  &
\colhead{Angle} &
\colhead{{H$\delta_{LS}$}} &
\colhead{{H$\gamma_{LS}$}} &
\colhead{{H$\beta_{LS}$}} }
\startdata
H1 & 46.7    &  13.63$^{+2.19}_{-2.29}$
             &  14.90$^{+1.04}_{-1.07}$
             &  10.35$^{+1.46}_{-1.49}$ \nl
& & & & \nl
H1 & 153.0   &  13.38$^{+3.38}_{-3.62}$ 
             &  10.06$^{+1.80}_{-1.88}$
             &  11.99$^{+2.35}_{-2.45}$ \nl
& & & & \nl
H1 Weighted Mean & \nodata
             &  13.56$^{+1.84}_{-1.94}$
             &  13.69$^{+0.90}_{-0.93}$
             &  10.80$^{+1.24}_{-1.27}$ \nl
& & & & \nl
\nl
H5 & 113.5   &  24.16$^{+4.04}_{-4.47}$
             &  19.58$^{+2.86}_{-3.15}$
             &  13.73$^{+7.49}_{-8.61}$  \nl
& & & & \nl
H5 & 140.0   &  20.78$^{+11.54}_{-15.47}$
             &   4.53$^{+8.67}_{-10.67}$
             &  -3.53$^{+17.61}_{-23.04}$ \nl
& & & & \nl
H5 Weighted Mean & \nodata & 23.85$^{+3.81}_{-4.29}$
             & 18.25$^{+2.71}_{-3.02}$
             & 11.39$^{+6.90}_{-8.06}$ \nl 
& & & & \nl
H4+9 & 113.5 &   4.91$^{+4.87}_{-5.29}$
             &  10.95$^{+2.06}_{-2.17}$
             &  -0.43$^{+4.06}_{-4.31}$ \nl
& & & & \nl
H2 &   113.5 &  19.08$^{+5.91}_{-6.75}$
             &   2.34$^{+3.23}_{-3.46}$
             & $(-36.43^{+8.30}_{-8.99})^*$  \nl 
& & & & \nl
H6 & 46.7    &  11.16$^{+7.88}_{-9.19}$
             &  12.55$^{+3.38}_{-3.71}$
             &  14.77$^{+4.21}_{-4.54}$  \nl
& & & & \nl
{\bf Arithmetic Mean} \nl
{\bf of All Except H2$^{**}$}& \nodata
             & {\bf $13.37\pm{3.94}$} 
             & {\bf $13.86\pm{1.56}$} 
             & {\bf $ 9.13\pm{3.30}$} \nl
\tablenotetext{*}{Reflects the presence of residual emission
from imperfect subtraction of the background emission line
spectrum of the galaxy.  Refer to Fig. 2.}
\tablenotetext{**}{H2 is excluded from the arithmetic mean because
of the above mentioned presence of residual Balmer line emission from
imperfect galaxy subtraction. The errors quoted for the mean
is the standard error of the mean.}
\enddata
\end{deluxetable}

\begin{deluxetable}{lrrrrcc}
\tablewidth{0pc}
\tablenum{5}
\tablecaption{Comparison of H1 EWs with Published Data}
\tablehead{
& \colhead{Position} & & & & & \\
\colhead{Source} &
\colhead{Angle}  &
\rcolhead{H$\delta_{BH}$} &
\rcolhead{H$\gamma_{BH}$} &
\rcolhead{H$\beta_{BH}$} &
\rcolhead{Mgb} &
\rcolhead{Fe52} \\
\colhead{(1)} &
\colhead{(2)} &
\colhead{(3)} &
\colhead{(4)} &
\colhead{(5)} &
\colhead{(6)} &
\colhead{(7)} }
\startdata
This study &  46.7 & 11.2  &   9.9  &  10.9  &   1.2  &  1.9 \cr
This study & 153.0 &  8.7  &   8.6  &  11.9  &   1.0  &  0.1 \cr
Zepf \etal &  0.0  & 13.3  &  10.7  &  14.7  &   0.6  &  2.6 \cr
Zepf \etal & 30.0  &  8.3  &  10.0  &  11.5  &   1.3  &  1.9 \cr
\tablenotetext{}{NOTE--All equivalent width values given in \AA.}
\tablenotetext{(1)}{Source of information}
\tablenotetext{(2)}{Counter-clockwise from N--S} 
\tablenotetext{(3)-(5)}{Balmer line EWs, as defined in Brodie \& Hanes (1986)} 
\tablenotetext{(6)-(7)}{As defined by Burstein \etal (1984)}
\enddata
\end{deluxetable}

\begin{deluxetable}{lrllllllcc}
\tablewidth{0pc}
\tablenum{6}
\tablecaption{Velocities of Clusters, Galaxy and Gas}
\tablehead{
& & & & \multicolumn{4}{c}{Gas velocity based on emission lines} \\ 
\cline{5-8}
\colhead{Cluster}    &
\colhead{Position}   &
\colhead{Cluster}    &
\colhead{Galaxy}     &
\colhead{H$\beta$ \hfil}  &
\multicolumn{2}{c}{[OIII]} &
\colhead{[OI]}  &
\colhead{Average} \\
\colhead{ID}         &
\colhead{Angle}      &
\colhead{Velocity}   &
\colhead{Velocity}   &
\colhead{4861\AA}  &
\colhead{4959\AA}  &
\colhead{5007\AA}  &
\colhead{6300\AA}  & 
\colhead{Gas Velocity} \\ 
\colhead{(1)} & \colhead{(2)} & \colhead{(3)} & \colhead{(4)} &
\colhead{(5)} & \colhead{(6)} & \colhead{(7)} & \colhead{(8)} &
\colhead{(9)} }
\startdata
H1   &	 46.7 &  5282$\pm$ 87 & 5221$\pm$ 71 &  5262  & 5370 & 5334 & 5250 & 5307$\pm$28  \nl
H1   &	153.0 &	 5111$\pm$ 82 & 5190$\pm$ 67 &  5251  & 5397 & 5388 & 5261 & 5327$\pm$39  \nl
H5   &	113.5 &  5293$\pm$146 & 5326$\pm$ 66 &  5367  & 5429 & 5370 & 5337 & 5379$\pm$18  \nl
H5   &	140.0 &  5150$\pm$149 & 5309$\pm$ 68 &  5346  & 5289 & 5349 & 5349 & 5336$\pm$15  \nl \nl
H1 &Wtd Mean  &  5191$\pm$ 60 & 5205$\pm$ 49 &  5257  & 5384 & 5361 & 5256 & 5314$\pm$33  \nl
H5 &Wtd Mean  &  5223$\pm$102 & 5318$\pm$ 47 &  5357  & 5359 & 5360 & 5343 & 5354$\pm$04 \nl
H6   &	 46.7 &  5208$\pm$110 & 5251$\pm$ 75 &  5221  & 5292 & 5232 & 5207 & 5241$\pm$18  \nl
H2   &  113.5 &	 5694$\pm$207 & 5343$\pm$ 98 &  5261  & 5174 & 5158 & 5275 & 5220$\pm$29  \nl
H4+9 &  113.5 &	 4553$\pm$104 & 5263$\pm$103 &  5244  & 5232 & 5223 & 5243 & 5239$\pm$04  \nl
\enddata
\tablenotetext{(2)}{Counter-clockwise from N--S}
\tablenotetext{(3)-(4)}{Measured by cross-correlation}
\tablenotetext{(5)-(8)}{Measured by measuring line centers}
\tablenotetext{(9)} {The arithmetic mean of the four line--of--sight
velocities measured for the gas.  Error reported is the standard error on
the mean.}
\end{deluxetable}

\begin{deluxetable}{lccccc}
\tablewidth{0pc}
\tablenum{7}
\tablecaption{Balmer Line Equivalent Widths of Young Star Clusters}
\tablehead{ &  \colhead{Estimated} & \colhead{Metallicity} & & \\
\colhead{Cluster ID} & \colhead{Age (Myr)} & \colhead{[Fe/H]} &
\colhead {H$\delta_{LS}$} & \colhead{H$\gamma_{LS}$} & \colhead{H$\beta_{LS}$} }
\startdata
NGC 1847/2157/2214 & 25      & -0.4  & 3.40  &  3.14  &  3.51 \nl
NGC 1866           & 80      & -0.5  & 7.50  &  6.48  &  6.96 \nl
NGC 1831/1868      & 200-500 & -0.6  & 9.27  &  8.34  &  8.13 \nl
\enddata
\tablenotetext{}{Spectra, ages and metallicities were obtained from
Bica \& Alloin (1986) as tabulated in the data base of Leitherer \etal(1996).}
\end{deluxetable}

\begin{deluxetable}{crrrrrr}
\tablewidth{0pc}
\tablenum{8}
\tablecaption{Fritze--v.Alvenzleben \& Kurth Spectral Synthesis Model Results} 
\tablehead{
\colhead{Age (Myrs)} & \colhead{B--V} & \colhead{V--R} & \colhead{B--R} &
\colhead{H$\beta$} (\AA) & \colhead{H$\gamma$ (\AA)} & \colhead{H$\delta$} (\AA) }
\startdata
 100 &  0.11  & 0.16 & 0.26 &  8.17 &  6.83 &  5.16 \nl
 150 &  0.12  & 0.16 & 0.28 &  9.60 &  8.08 &  6.06 \nl
 200 &  0.14  & 0.15 & 0.29 & 10.52 &  8.47 &  6.61 \nl
 250 &  0.17  & 0.16 & 0.33 & 11.14 &  8.77 &  6.95 \nl	
 300 &  0.20  & 0.18 & 0.38 & 11.33 &  8.84 &  7.02 \nl
 350 &  0.24  & 0.20 & 0.44 & 11.35 &  8.81 &  6.98 \nl
 400 &  0.27  & 0.21 & 0.48 & 11.05 &  8.55 &  6.67 \nl
 450 &  0.30  & 0.22 & 0.52 & 10.73 &  8.24 &  6.30 \nl
 500 &  0.33  & 0.24 & 0.57 & 10.37 &  7.92 &  6.05 \nl
\nl			     
H1  &        &      & 0.56--0.24 & 11.12 &  8.68 &  9.21 \nl
\enddata
\tablenotetext{NOTES}
\tablenotetext{}
\tablenotetext{}{Note that the equivalent width values
listed here were measured on the model spectra
using customized bandpass definitions that are {\it narrower}
than the LS indices and therefore are comparable only to each other,
not to the values listed in Tables 4 and 7. \nl}
\tablenotetext{}{
The color range 
for H1 is defined by the color uncorrected for internal reddening
and the color corrected for the maximum likely extinction for
this cluster. See section 3.}

\end{deluxetable}

\begin{deluxetable}{lllllll}
\tablewidth{0pc}
\tablenum{9}
\tablecaption{Comparison of Measured Metal Absorption Line Strengths with Model$^{*}$ Spectra of 350 Myr Old Clusters of Various Metallicities} 
\tablehead{
\colhead{z} & \colhead{Mg2$^{**}$} & \colhead{Mg1$^{**}$} & \colhead{Mgb} & \colhead{Fe5335} & \colhead{Fe5270} & \colhead{NaI}}
\startdata

0.008 & 0.043  &  0.017  & 0.67  & 0.66  & 0.54  & 0.48  \nl
0.020 & 0.056  &  0.023  & 0.81  & 0.77  & 0.68  & 0.68  \nl
0.040 & 0.079  &  0.035  & 1.11  & 1.07  & 1.01  & 1.11  \nl
\nl
H1 Data   & 0.001   & -0.016  &  1.14  &  0.58   & 1.00   & 5.38  \nl
Avg Error & 0.026   &  0.025  &  0.77  &  1.42   & 0.76   & 0.43  \nl
\enddata
\tablenotetext{*}{From the spectral synthesis models of Fritze-v.Alvensleben \& Kurth (1997).}
\tablenotetext{**}{Mg2 and Mg1 are given in magnitudes.  All other line strengths are given in \AA.}
\end{deluxetable}

\onecolumn 

\clearpage

\clearpage

\clearpage

\clearpage

\clearpage

\clearpage

\clearpage

\end{document}